\documentclass[]{aa} 
\usepackage{graphicx}
\usepackage{txfonts}
\usepackage{color}
\usepackage{natbib}
\usepackage[british]{babel}
\usepackage{setspace}
\usepackage{longtable}
\usepackage{url}

\begin{document}
   \title{Full spectral fitting of Milky Way and M31 globular clusters: ages and
   metallicities}

   \subtitle{}

   \author{Enio Cezario
          \inst{1}
          \and
          Paula R. T. Coelho
          \inst{1}
          \and
          Alan Alves-Brito
          \inst{2}
          \and
          Duncan A. Forbes
          \inst{3}
          \and
          Jean P. Brodie
          \inst{4}
          }

   \institute{N\'ucleo de Astrof\'{\i}sica Te\'orica, Universidade Cruzeiro do Sul, 
R. Galv\~ao Bueno 868, 01506-000, S\~ao Paulo, Brazil\\
              \email{paula.coelho@cruzeirodosul.edu.br}
         \and
             {Research School of Astronomy and Astrophysics,
The Australian National University, Cotter Road, Weston, ACT 2611, Australia}\\
             \email{abrito@mso.anu.edu.au}
         \and
             {Centre for Astrophysics and Supercomputing, Swinburne University of
Technology, Hawthorn, Victoria 3122, Australia}\\
             \email{dforbes@swin.edu.au}
         \and
             {UCO/Lick Observatory, University of California, Santa Cruz, CA 95064,
USA}\\
             \email{brodie@ucolick.org}
             }

   \date{}
   
   \authorrunning{Cezario et al.}
   \titlerunning{Spectral fitting of globular clusters}

 
  \abstract
   {The formation and evolution of disk galaxies are long standing questions in Astronomy. Understanding the properties of globular cluster
   systems can lead to important insights on the evolution of its host galaxy.}
   {We aim to obtain the stellar population parameters -- age and metallicity -- of a sample of M31 and Galactic globular clusters. Studying their globular
   cluster systems is an important step towards understanding their formation and evolution in a complete way. }
   {Our analysis employs a modern pixel-to-pixel spectral fitting technique to fit observed integrated spectra to updated stellar 
   population models. By comparing observations to models we obtain the ages and metallicities of their stellar populations. We apply 
   this technique to a sample of 38 globular clusters in M31 and to 41 Galactic globular clusters, used as a control sample.}
   {Our sample of M31 globular clusters spans ages from 150~Myr to the age of the Universe. Metallicities [Fe/H] range from 
    --2.2 dex to the solar value. The age-metallicity relation obtained can be described as having two components: 
an old population with a flat age-[Fe/H] relation, possibly associated with the halo and/or bulge, and a second one with a roughly linear relation between age and metallicity, higher metallicities corresponding to younger ages,
possibly associated with the M31 disk. While we recover the very well known Galactic GC
metallicity bimodality, our own analysis of M31's metallicity distribution function (MDF) suggests that both
GC systems cover basically the same [Fe/H] range yet M31's MDF is not clearly bimodal. These results
suggest that both galaxies experienced different star formation and accretion histories. }
   {}


   \maketitle
%

\section{Introduction}

Globular clusters (GCs) are widely considered as excellent astrophysical
laboratories. Their age and metallicities, in particular, trace the
main (astro)physical processes responsible for the formation and evolution of
their host galaxies. However, differently from the Milky Way,
most of the extragalactic GCs cannot be resolved into individual stars. 

The techniques to estimate ages and metallicities of extragalactic
GCs have classically fallen into two broad categories. Those relying on photometry
\citep[e.g.][]{fan+06} are more susceptible to the age-metallicity degeneracy in the sense that young metal-rich populations are photometrically
indistinguishable from older metal-poor populations \citep[but see, e.g.,][]{ma+07}. The different
spectroscopic methods, however, are inspired
by the Lick/IDS system of absorption line indices \citep[e.g.][and references
therein]{worthey+94} --- from linear
metallicity calibrations \citep[e.g.][]{brodie_huchra90} to those which perform the
simultaneous $\chi$$^{2}$-minimisation of a large number of 
spectral indices \citep[see, e.g.,][]{proctor+04}. Nevertheless, both photometric and spectroscopic methods are dependent on the accurate
modelling of simple stellar
populations
\citep[SSPs, see, e.g.,][]{BC03,PEGASE-HR,delgado+05,maraston05,coelho+07,percival+09,lee+09,conroy_gunn10,vazdekis+10}.  

More recently, it is becoming  increasingly common to use spectral fitting on a pixel-to-pixel basis
to study integrated spectra of stellar clusters \citep[e.g.][]{dias+10}. This technique is an 
improvement over older methods
and has been recently discussed in e.g. \citet{koleva+08,cid_delgado10} and references therein.
This method has advantages such as 
making use of all the information available in a spectrum (making it possible to perform analysis at lower S/N) and not being limited by the physical broadening, since the internal kinematics is determined simultaneously with the population parameters. In some of its flavours, this method is also insensitive to extinction or flux calibrating errors. 
It has also been shown that full spectrum fitting reproduces better
the results from colour-magnitude diagrams (CMDs) than other methods \citep[e.g.][]{wolf+07}.

In this study, we have used the spectrum fitting code {\it ULySS}\footnote{http://ulyss.univ-lyon1.fr}  \citep{ulyss}
to compare, on a pixel-by-pixel basis, the integrated spectrum of 38 spectra of
GCs in M31 to simple stellar population models in
order to derive their ages and metallicities. 
Our sample comprises 35 integrated spectra previously analysed by \citet{beasley+04}, and
three outer halo clusters taken from \citet{abrito+09}.
In addition, we have also analysed integrated spectra of 41 Galactic GCs
presented in \citet{schiavon+05}, whose populations have been studied with CMDs and spectroscopy of individual stars.
Both, Galactic and M31 GCs were analysed in the same way. The Galactic
sample acts
as a control sample to estimate the 
reliability of the fitting method. 

Our nearest \citep[$\sim$780
kpc,][]{holland98} giant spiral galaxy M31 is a natural
target to test the current ideas about the formation and evolution of galaxies
in the Local Universe. As pointed out in \citet{abrito+09}, a 
remarkable difference between the M31 and the Galactic GC system is
that M31 hosts more GCs (by a factor of 2--3) than the Milky Way. 
Furthermore, there have been suggestions that M31 may have a significant population of young and intermediate 
age (less than 8 Gyr) that is not found in the Galactic 
system. The metallicity distribution 
function (MDF) of M31 (through both giant stars and 
GCs) and other galaxies \citep[see, e.g.][]{abrito+11,usher+12} has been a controversial topic regarding
its shape and distribution when
compared with our own Galaxy. In particular, the presence (or not) of colour-[Fe/H] bimodality is one
of the interesting and intriguing open questions in the field. Therefore, these
(dis)similarities between the different spiral galaxies in the Local Group  need to be investigated
through different techniques to better
understand the different process(es) in which spiral galaxies are formed and
have evolved in the Universe.

The paper is organised as follows. In Sect. 2 we describe our sample.
In Sect. 3 we present the stellar population analysis.
In Sect. 4 we discuss the results obtained. 
Concluding remarks are given in Sect. 5.


\section{Sample}

\subsection{M31 GCs}

The integrated spectra used in this study were previously studied by
\citet{abrito+09} and \citet{beasley+04}. 
The former provided spectra for three
outer halo GCs at projected distances beyond $\approx$ 80 kpc from M31. These
spectra were observed with the cross-dispersed, high-resolution spectrograph
HIRES instrument on the Keck 1 telescope, covering a broad wavelength range of
$\lambda\lambda=4020$--$8520$\,\AA, at a spectral resolving power of
$R\approx50,000$. Spectroscopic age and metallicities were obtained by using
metallicity calibrations from Mg{\it b}, CH and Mg2 indices. In addition, the
authors also employed the
simultaneous $\chi$$^{2}$-minimisation of a large number of 
spectral indices, approach introduced by \citet{proctor+04}.

The second source of data is detailed in \citet{beasley+04}, who made a
 analysis of high-quality integrated spectral indices in M31. We have
only used the spectra obtained with the Low Resolution Imaging Spectrograph mounted
on the Keck I telescope, covering a spectral range of
$\lambda\lambda=3670$--$6200$\,\AA\ and with a full width at half-maximum (FWHM)
resolution of 5\,\AA. The sample chemical properties were studied 
through the measurement of Lick line strengths.

In Table \ref{tab_m31} we list the M31 GCs analysed in the present work, and
we refer the reader to \citet{abrito+09} and \citet{beasley+04} for additional
information about the observations and data reduction.
The projected distances shown in Table \ref{tab_m31} were calculated
relative to an adopted M31 central position of $\alpha_{\rm J2000}$ = 00$^{\rm h}$42$^{\rm
m}$44$^{\rm s}$.30, $\delta_{\rm J2000}$ = +41$^{\rm o}$16$^{\prime}$09.90\farcs4.
In addition, a position angle for the X-coordinate of 38$^{\rm o}$
\citep{kent+89} as well as a distance of 780 kpc \citep{holland98} were
adopted. At this distance, 1 arc-min corresponds to 228 pc.

\begin{table*}
\caption{M31 GCs analysed in the present work. 
}
\label{tab_m31}
\centering
\begin{tabular}{ l c c r r c }
\hline
    ID    &    R.A. (J2000.0) &   Decl. (J2000.0) &  d$_p$ & S/N &  Reference \\
          &                   &                   &  (kpc)      & (pixel$^{-1}$) & \\
\hline    
    B126   & 00:42:43.481   & +41:12:42.47    &    0.79         &73   &  a\\
    B134   & 00:42:51.678   & +41:44:03.42    &    0.52         &70   &  a\\
    B158   & 00:43:14.406   & +41:47:21.28    &    2.40         &120  &  a\\
    B163   & 00:43:17.640   & +41:27:44.91    &    3.01         &170  &  a\\
    B222   & 00:44:25.380   & +41:14:11.62    &    4.37         &30   &  a\\   
    B225   & 00:44:29.560   & +41:21:35.27    &    4.69         &169  &  a\\
    B234   & 00:44:46.375   & +41:29:17.77    &    6.03         &71   &  a\\
    B292   & 00:36:16.666   & +40:58:26.58    &    17.15        &32   &  a\\
    B301   & 00:38:21.581   & +40:03:37.16    &    20.09        &31   &  a\\
    B302   & 00:38:33.500   & +41:20:52.29    &    10.81        &28   &  a\\  
    B304   & 00:38:56.940   & +41:10:28.41    &    9.85         &34   &  a\\
    B305   & 00:38:58.922   & +40:16:31.78    &    16.74        &15   &  a\\    
    B307   & 00:39:18.477   & +40:32:58.05    &    13.27        & 20  &  a\\
    B310   & 00:39:25.752   & +41:23:33.14    &    8.67         &35   &  a\\
    B313   & 00:39:44.599   & +40:52:55.05    &    9.38         &52   &  a\\
    B314   & 00:39:44.599   & +40:14:07.95    &    16.15        &17   &  a\\
    B316   & 00:39:53.604   & +40:41:39.29    &    10.78        & 20  &  a\\
    B321   & 00:40:15.545   & +40:27:46.50    &    12.78        &30   &  a\\
    B322   & 00:40:17.270   & +40:39:04.70    &    10.57        &69   &  a\\
    B324   & 00:40:20.477   & +41:40:49.38    &    8.33         &71   &  a\\
    B327   & 00:40:24.107   & +40:36:22.52    &    10.91      &30   &  a\\
    B328   & 00:40:24.529   & +41:40:23.15    &    8.13       &30   &  a\\
    B331   & 00:40:26.642   & +41:42:04.28    &    8.35       &15   &  a\\
    B337   & 00:40:48.477   & +42:12:11.06    &    13.71       &178  &  a\\
    B347   & 00:42:22.892   & +41:24:27.58    &    8.79       &61   &  a\\
    B350   & 00:42:28.442   & +40:24:51.12    &    11.73       &39   &  a\\
    B354   & 00:42:47.645   & +42:00:24.73    &    10.10          &30   &  a\\
    B365   & 00:44:36.445   & +42:17:20.57    &    14.77      &60   &  a\\
    B380   & 00:46:06.238   & +42:00:53.13    &    13.36      &43   &  a\\
    B383   & 00:46:11.498   & +41:19:41.48    &    8.94       &85   &  a\\
    B393   & 00:47:01.204   & +41:24:66.39    &    11.16      &43   &  a\\
    B398   & 00:47:57.786   & +41:48:45.66    &    15.32      &44   &  a\\
    B401   & 00:48:08.508   & +41:40:41.93    &    14.96      &54   &  a\\
    MGC1   & 00:50:42.459   & +32:54:58.78    &    116.05        & 20  &  b\\
    MCGC5  & 00:35:59.700   & +35:11:03.60    &    78.49         & 20  &  b\\
    MCGC10 & 01:07:26.318   & +35:46:48.41    &    99.85         & 20  &  b\\
    NB16   & 00:42:33.094   & +41:20:16.41    &    1.05          &142  &  a\\
    NB89   & 00:42:44.780   & +41:14:44.20    &    0.34          &110  &  a\\
\hline\end{tabular}    
\begin{minipage}{.88\hsize}
{\bf Notes.---} The first column gives the GC identification, as in the Bologna
catalogue \citep{galleti+04}. Second and third columns give the coordinates of the objects, in
J2000.
Fourth column shows the projected distance of the cluster to the centre of M31 (see text in \S 2.1). 
Fifth column shows the S/N of the spectra as given in the corresponding paper; when this information was not available  (clusters B302, B305, B307, B316, B331, B354),
we estimated the S/N around 5000\,{\AA} using standard IRAF\footnote{IRAF is distributed by the National Optical Astronomy
Observatory, which is operated by the Association of Universities for Research in Astronomy, Inc., under cooperative
agreement with the National Science Foundation.} routines.
Sixth column gives the reference of the spectra: (a) for \citet{beasley+04} and
(b) for \citet{abrito+09}.
\end{minipage}    
\end{table*}

\subsection{Galactic GCs}
The sample of Galactic GCs spectra is given in Table \ref{tab_res_mw}, whose observations
were taken from 
\citet{schiavon+05}. 
The observations were performed with a long-slit spectrograph in drift-scan mode 
in order to integrate the population within one core radius. The spectra cover 
the range $\lambda\lambda$ = 3350 -- 6430 $\AA$ at a resolution of about 
FWHM = 3\,$\AA$. The mean S/N varies from 50 to 240 $\AA^{-1}$ depending on the
wavelength. We refer the reader to \citet{schiavon+05} for more details on the observations and data reduction.

The aim of analysing this Galactic sample is twofold: first, we use it as a control 
sample where we can evaluate
the performance of the fitting technique by comparing our results to studies
of CMD and spectroscopy of individual stars; and second, by ensuring that the same analysis is applied
consistently to the M31 and Galactic sample, the latter acts as a reference system to
which the M31 GC system is compared. 

We searched the literature for good independent determinations of ages and metallicities of this
sample. Fitting theoretical isochrones to GC CMDs is generally accepted as the most secure age
determination possible when using photometry; however, the results do vary between sets of isochrones
and methods of analysis, and the absolute derived age depends on model zero points, input physics, colour-T$_{eff}$ transformation, distance uncertainties and foreground reddening
\citep[e.g.][]{chaboyer+98,buonanno+98,meissner_weiss06}. In this work we relied on the
relative ages derived homogeneously for a large sample of clusters from
\citet{deangeli+05} and \citet{marin-franch+09}. Those relative ages were converted to
absolute ages by adopting a conservative value of 13 $\pm$ 2.5 Gyr for 47
Tucanae \citep{zoccali+01}. We additionally added values for NGC~6528 and NGC~6553 from \citet{momany+03} and \citet{beaulieu+01}, respectively. 

Regarding the metallicities, we adopt two homogeneous
compilations found in literature: the compilation by \citet{carretta+09}, who analysed high resolution stellar spectra in 19 globular clusters, and 
brought the lower resolution measurements from \citet{zinn_west84,kraft_ivans03,rutledge+97} to a
common metallicity-scale; and the compilation by \citet{schiavon+05}, based on measurements from
\citet{kraft_ivans03, carreta_gratton97}.


\section{Analysis}

We obtained ages and metallicities for the GCs
through the comparison of their integrated spectra to SSP models, using the public code \emph{ULySS} \citep{ulyss}, described briefly below. 

\emph{ULySS}
  is a software package
 performing spectral fitting in two astrophysical contexts: 
the determination of stellar atmospheric parameters and the study
of the star formation and chemical enrichment history of galaxies. 
In {\it ULySS}, an observed spectrum is fitted against a model 
(expressed as a linear combination of components)
through a non-linear least-squares minimisation.
In the case of our study, the components are SSP models. 

We adopt SSP models by \citet{vazdekis+10}, which
cover the wavelength range 3540\,--\,7400\AA\, at a resolution of FWHM $\sim$ 2.5\AA.
Ages range between 63\,Myr and 18\,Gyr, and metallicities [Fe/H] between --2.32
and +0.22 dex. These models are based on
MILES stellar library \citep{MILES1,MILES2} and \citet{girardi+00} evolutionary tracks. Being based
on an empirical library, the models follow the abundance
pattern of the solar neighbourhood, namely they are solar-scaled around solar metallicities, 
and $\alpha$-enhanced for low metallicities \citep{milone+11}.

{\it ULySS} matches model and
observation continua through a multiplicative polynomial, 
determined during the fitting process. Therefore, {\it ULySS}
is not sensitive to flux calibration, galactic extinction, or any other
cause affecting the shape  of the spectrum. We run  
{\it ULySS} with its global minimisation option, i.e., each fitting was performed starting 
from several guesses; this is an important feature that minimises the risks of results being biased
by local minima. 

For each spectrum we run 300 Monte Carlo (MC) simulations.
The simulations consist of analyses of the spectrum with a random noise added, according to the S/N of the observation.
As explained in \citet{ulyss}, the MC simulations take into account the correlation between 
pixels and thus reproduce the correct noise spectrum, giving a robust estimate of the errors.
We adopt the final stellar populations parameters to be the 
mean values of the 300 simulations, and our final uncertainties are the 1\,$\sigma$
value of the simulations results. We quote a minimum error of 10\% in age to account for possible systematic errors in the isochrones used in the SSP models \citep[e.g.][]{gallart+05},
when the 1\,$\sigma$ value is smaller than this limit. 

A possible disadvantage of spectral fitting, compared to Lick indices, is that 
the method may in principle be sensitive to the wavelength range fitted. Reports on
this effect in literature are  contradictory: \citet{koleva+08} has 
concluded that the sensitivity to the wavelength range is not critical; at odds
with this conclusion, \citet{walcher+09} reports that while fitting globular
clusters of the Galactic bulge, the results are more accurate when
the fitting is limited to a wavelength range $\sim$~530\,{\AA} wide centred at 5100\,{\AA}. 

We thus performed some tests on the dependence with the wavelength window by
fitting the Galactic GCs sample at different wavelength ranges. The widest range we tested was 
3650 -- 6150 {\AA} (the total range covered by \citealt{beasley+04} observations) and the shortest 4828 -- 5364 {\AA} 
(favoured by \citealt{walcher+09}). We also tested the wavelength used in 
\citet[][4000 -- 5700 {\AA}]{koleva+08} and the range 4000 -- 5400 {\AA}, similar to the coverage of \citet{abrito+09} observations.


\section{Results}

\subsection{\label{galgc}Stellar populations in the Galaxy --- verifying the method}

By analysing the Galactic sample at different wavelength windows, we verified that 
the wavelength choice has little influence on the
metallicities derived via spectral fitting. 
On the other hand, the ages may change in a non-negligible way: in particular, fitting the widest
range resulted in nearly a third of the GCs being fitted with intermediate ages (down to 4\,Gyr). 
Possible sources of errors are: blue horizontal branches or blue stragglers not properly taken into account on the models, which is known to affect spectroscopic
ages both in spectral fitting and
Lick indices \citep[e.g.][]{pacheco_barbuy95,koleva+08}, deficiencies in the observations (diffuse light affecting the blue region and poor subtraction of telluric lines affecting the red, as mentioned in \citealt{koleva+08}) and limitations in the SSP models (chemical patterns different than the ones in MILES library). 

\begin{figure}[h]
\begin{center}
\includegraphics[width=1.0\columnwidth]{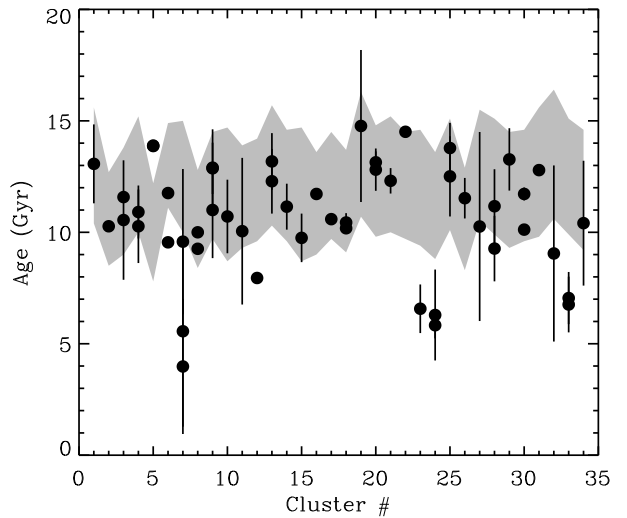}
\caption{\label{fig_diff_age_ours}Comparison between the ages derived via spectral fitting (filled
circles) and ages derived via CMD (filled grey pattern) for our sample of Galactic GCs. The number in the $x$-axis identifies the GCs
{\tiny (1: NGC~104;
2: NGC~1851;
3: NGC~1904;     
4: NGC~2298;
5: NGC~2808;
6: NGC~3201;
7: NGC~5286;
8: NGC~5904;
9: NGC~5927;
10: NGC~5946;
11: NGC~5986;
12: NGC~6121;
13: NGC~6171;
14: NGC~6218;
15: NGC~6235;
16: NGC~6254;
17: NGC~6266;
18: NGC~6284;
19: NGC~6304;
20: NGC~6342;
21: NGC~6352;
22: NGC~6362;
23: NGC~6388;
24: NGC~6441;
25: NGC~6528;
26: NGC~6544;
27: NGC~6553;
28: NGC~6624;
29: NGC~6637;
30: NGC~6652;
31: NGC~6723;
32: NGC~6752;
33: NGC~7078;
34: NGC~7089).}
}
\end{center}
\end{figure}

\begin{figure}[ht]
\begin{center}
\includegraphics[width=1.0\columnwidth]{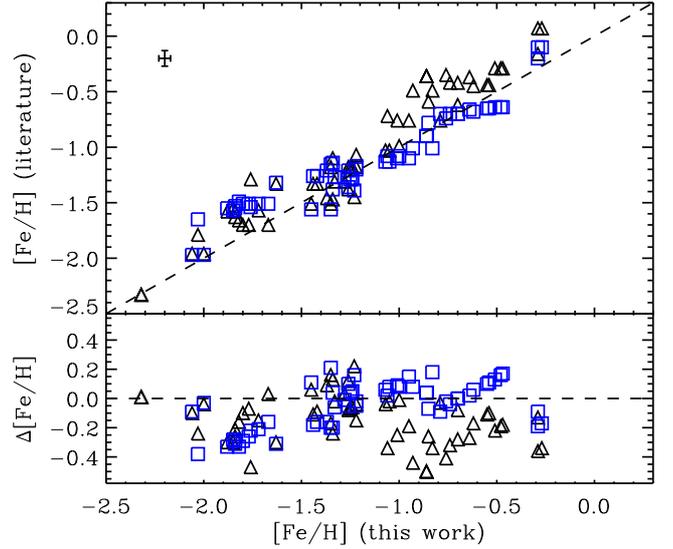}
\caption{\label{fig_diff_fe_ours} {\it Top panel}: Comparison of metallicities obtained with the spectral fitting (this
   work) versus the values compiled in literature by \citet{schiavon+05} (blue squares) and \citet{carretta+09} (black triangles) for
   the Galactic GCs. The error bar in the top left corner indicates the mean error values. {\it Bottom panel}: [Fe/H] residuals (ours minus literature values).}
\end{center}
\end{figure}

In the present work we favour the wavelength range 
4000 -- 5400{\AA} which, among the wavelength windows we tested, is the 
one which best reproduced the CMD ages of the Galactic globular 
clusters. The results of this fitting run are shown in Table \ref{tab_res_mw} together with literature values. Comparisons between our results and literature are also shown in Figures \ref{fig_diff_age_ours} and \ref{fig_diff_fe_ours} for ages and metallicities, respectively. Several of the Galactic GCs have multiple
observations and we fitted each spectrum 
individually.

In Fig. \ref{fig_diff_age_ours} it can be seen that  six clusters have spectroscopic ages which are outside the range of  
allowed CMD values:  NGC 2808 (GC \#5 in Fig. 1), NGC 5286 (GC \#7), NGC 6121 (GC \#12), NGC 6388 (GC \#23), NGC 6441 (GC \#24), NGC 7078 (GC \#33).

In the case of NGC 5286 (GC \#7), it is striking that three observations of the same
cluster resulted in ages different by almost 6~Gyr. According to the observation log
in \citet{schiavon+05}, the three observations of this GC were taken at
different slit positions or extraction aperture. A visual inspection of
the fitting showed that in two exposures the observed Balmer lines are visibly narrower
than the model, even though the metallic lines are well fitted (this pattern is also present in NGC 6752, GC \# 32). We suspect that  
contamination from foreground stars hampered the observations. A visual inspection of the field around this GC in \emph{Aladin}\footnote{\url{http://aladin.u-strasbg.fr}} shows a handful of bright foreground stars close to the cluster. Given that the observations were performed in
drift-scan mode, diffuse light from one of these stars might explain the failure of fitting the line profiles. For the remaining clusters, there was no obvious pattern in the visual inspection of the fitting, so we further investigated other sources of problems that could impact the derived ages.

It is well established in literature \citep[e.g.][]{gratton+12} that virtually all clusters harbour populations with anti-correlated C--N, O--Na (and sometimes Mg--Al) abundances. In this sense, a comparison to strict SSP models might not be adequate in many cases. In \citet{coelho+11,coelho+12proc} the authors investigate, by means of stellar population modelling, the effect of populations with CNONa variations in spectral indices and integrated spectra. Inside the wavelength range fitted in this work, we obtain from Fig. 5 in \citet{coelho+12proc} that the 
range $\sim$4150--4220{\AA} is  affected by CNONa anti-correlated abundances. We investigated if these chemical variations could explain the age mismatch by repeating the fitting of the clusters NGC 2808, NGC 6121, NGC 6388, NGC 6441, NGC 7078, masking the CNONa affected region. We
verified that the masking does not improve the age results, the differences between masking or not the CNONa region being $\Delta$age = 0.1 $\pm$ 0.2 Gyr.
We conclude therefore that CNONa anti-correlations in the integrated spectra of clusters cannot explain the cases where 
spectroscopic ages do not match the CMD values. 

The test above does not guarantee against the effect of multiple main sequences and/or sub-giant branches, clusters for which remarkably a single age isochrone is not 
adequate to fit the CMD. This is the case of  
NGC~2808 \citep{piotto+07}, which is known to harbour three main sequences and a very complex horizontal branch. In such striking cases, it is not disquieting that comparing the observations to SSP models would fail.

Except for NGC~2808, the other clusters with deviating spectroscopic ages are all younger than the CMD ages by at least 2--3 Gyr. Our initial interpretation was 
the potential presence of HB morphologies which are not well represented in the SSP models. 
Extended HB morphologies are long known in literature to bias spectroscopic ages towards lower values \citep[e.g.][]{pacheco_barbuy95,lee+00,schiavon+04b,mendel+07,ocvirk10}. \citet{koleva+08} manages to reconcile some of the spectroscopic ages of galactic clusters with CMD measurements when a hot star component is added to the fitting, to mimic the presence
of blue horizontal branch stars or blue stragglers. It is not straightforward, however, to predict how the
HB morphology will impact the spectroscopic ages. The effective impact of the HB morphology 
on spectroscopic age is likely a non-trivial interplay 
between the wavelength range fitted (in the sense that bluer regions will be more 
sensitive to hotter HB stars) and the exact morphology (or lack of) predicted by
the underlying isochrone of the stellar population model, this morphology also being dependent 
on the age and metallicity of the modelled population.

Using the study by \citet{gratton+10a} on the galactic clusters HB
morphologies, we searched for patterns of the HB morphologies that could
correlate with the clusters with deviant ages, but could not find any. The only
note of interest is that four of these clusters (NGC2808, NGC6388, NGC6441, NGC7078) have high values of 
R'\footnote{R' = N$_{HB}$/N'$_{RGB}$, where N$_{HB}$ is the number of stars in the HB and N'$_{RGB}$ is the number of stars on the RGB brighter than V(HB)+1.} (larger than 0.8). Nevertheless, cluster NGC~5927 (GC \#9) also has a very high R' and its spectroscopic age matches the CMD range. 

Regarding the metallicities derived via spectral fitting, in Fig. \ref{fig_diff_fe_ours} we compare our [Fe/H] values with those compiled by
\citet{carretta+09} (black triangles) and \citet{schiavon+05} (blue squares). 
The overall
difference (this work $-$ Schiavon et al.) is of --0.05 $\pm$ 0.16 dex,
while (this work $-$ Carretta et al.) is --0.15 $\pm$ 0.17 dex.
Even though the agreement with Schiavon et al. scale is better on average, it was pointed out by the referee that there
seems to be a slope between our results and Schiavon et al. scale (the relation crossing the line of equality). On the other hand, the offset between our results and Carretta et 
al. scale is nearly constant, thus the differences between our results and this scale might be explained by a zero-point offset.

The
scatter we obtain between our metallicities and the stellar analysis is compatible with the difference between Schiavon et al. and Carretta et al. scales 
(0.10 $\pm$ 0.20 dex), which are both based on high resolution stellar
spectroscopy. It is a remarkable agreement between metallicities from {\it integrated light at medium
spectral resolutions} and high-resolution stellar analysis.

From the analysis of the galactic sample we conclude that regarding the age determination, the method returned accurate
results except for 
six clusters: NGC 2808, NGC 5286, NGC 6121, NGC 6388, NGC 6441, NGC 7078.
In the case of NGC~5286 (and possibly NGC~6752, whose spectroscopic age
marginally matches the CMD range), we suspect that contamination from foreground
stars hampered the observations. NGC~2808 is a striking case of a cluster with
triple main sequences, and a failure when comparing to SSP models is not
disquieting. For the remaining four clusters, or $\sim$12\% of our galactic sample, ages are underestimated by $\sim$2--3 Gyr and we cannot provide a robust explanation for these differences. We confirmed that the presence of populations with CNONa anti-correlated abundances cannot explain the age differences. If this age difference is related to HB morphologies, as usually claimed in literature, its exact effect remains to be better understood, given that not {\it all} GC with extended HB were affected. We  reproduced the CMD ages of other GCs known to have blue components (such as
NGC~5946 and NGC~6284), without invoking additional parameters in the fitting
(such as a free amount of hot stars in \citealt{koleva+08}). The mean
difference (this work - literature) is of --0.8 Gyr with an r.m.s. of 1.7 Gyr,
for the sub-sample with fitted ages inside the observational uncertainties; and
--1.8 Gyr with a r.m.s. of 2.8 Gyr for the whole sample.

As for the metallicities, the method gives results in accordance with
determinations from high resolution stellar spectroscopy (R$\large$ 30,000), with a r.m.s. of the
same order of the r.m.s. between two different sets of high-resolution results
as presented in the previous paragraph.

\addtocounter{table}{1}

\subsection{Stellar populations in M31 GCs}
\begin{figure*}
\begin{center}
\begin{tabular}{cc}
\includegraphics[height=6cm,trim=0.3cm 0 0 0.8cm,clip=true]{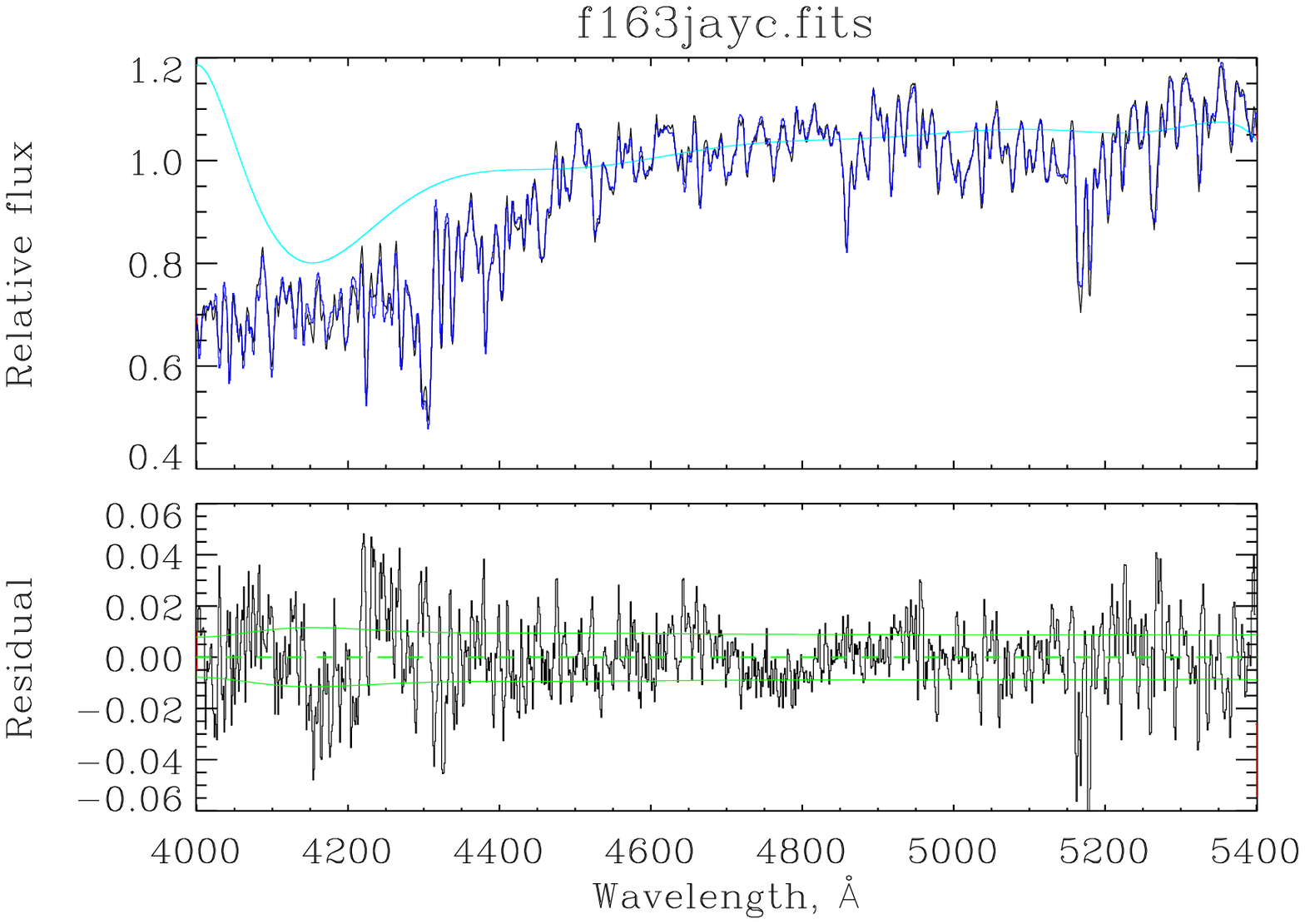}
\includegraphics[height=6cm,trim=1.2cm 0 0 0.8cm,clip=true]{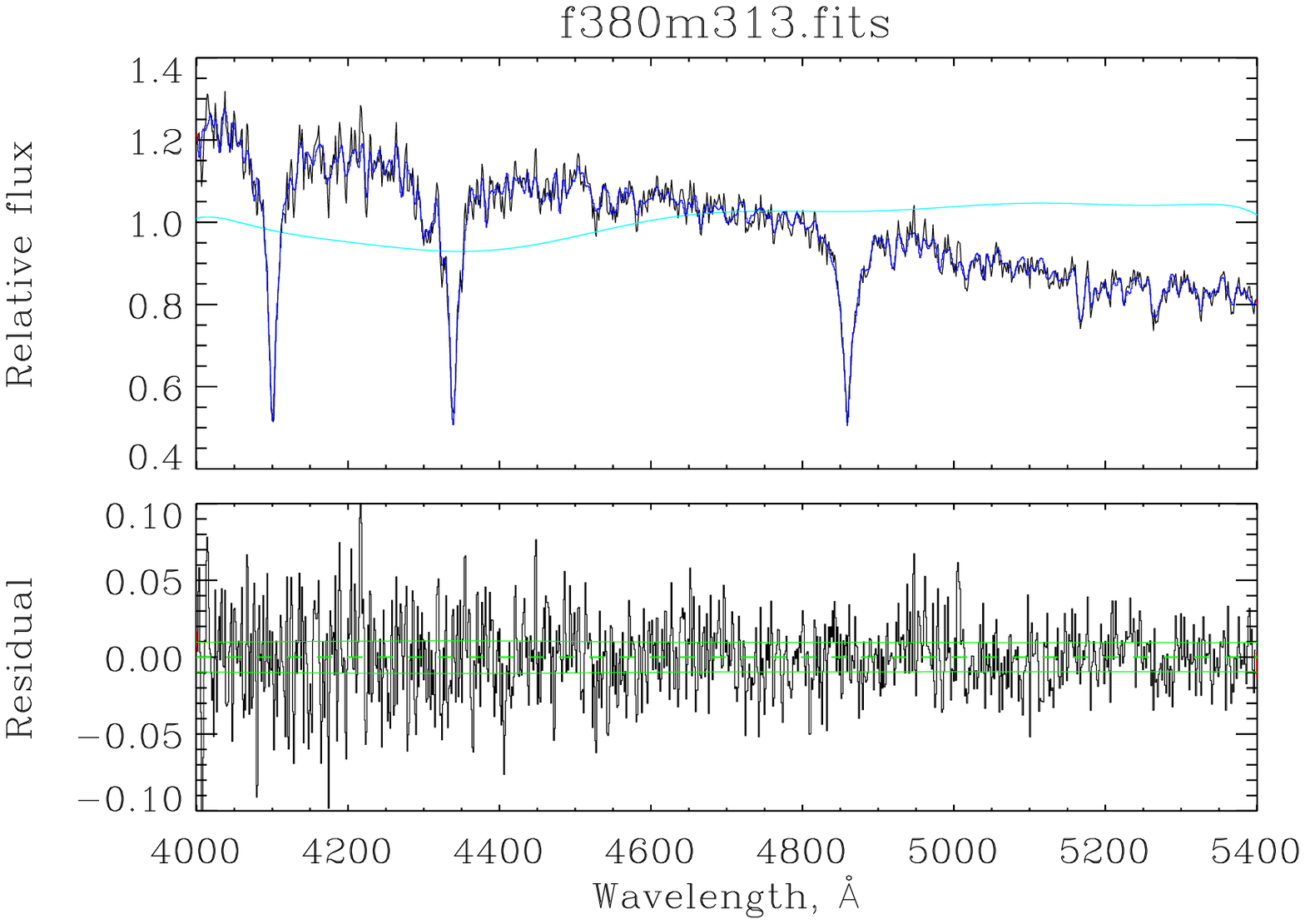}
 \end{tabular}
\caption{\label{f:fit} Spectral fitting of the GCs B163 (left-hand panel) and B380 (right-hand panel). The top panels show the spectra in black and the best fit in blue. The cyan lines are the multiplicative polynomials. The bottom panels are the residuals from the best fit, where
the continuous green lines mark the 1-$\sigma$ deviation as expected by the input S/N, and the dashed green lines mark the zero y-axis. This plot is best seen in colour (online version only).}
\end{center}
\end{figure*}

We analysed the M31 sample with the same set up and procedure as in the galactic sample, and our derived ages and metallicities are presented in Table \ref{tab_res_m31}.
We show in Fig. \ref{f:fit} the fitting of two spectra in our M31 sample, for illustration purposes.
We obtain a large range of ages, from $\sim$150~Myr (B322) to very old ages. 
In fact, three of the objects (B163, B393, and B398) were given ``older than the Universe" ages \citep[13.75 $\pm$ 0.11 Gyr;][]{jarosik+11}. In these cases, $\chi^2$ 
maps of age distributions show a valley of low values starting around $\sim$~6\,Gyr
and almost flat with metallicity, indicating that the results are degenerate in age. The metallicities in our sample range from --2.2 dex to
+0.1 dex. 
 
\begin{table}[h]
\caption{Ages and metallicities for M31 GCs derived in this work.}
\label{tab_res_m31}
\centering
\begin{tabular}{lrr}
\hline
    Cluster    &    Age (Gyr)     &   [Fe/H]      \\
\hline
B126       &    2.41 $\pm$    0.86 &   -1.19 $\pm$    0.15 \\
B134       &   13.72 $\pm$    2.70 &   -0.89 $\pm$    0.04 \\
B158       &   12.44 $\pm$    1.24 &   -0.74 $\pm$    0.02 \\ 
B163       &   16.86 $\pm$    0.43 &   -0.29 $\pm$    0.01 \\
B222       &    1.16 $\pm$    0.12 &   -0.28 $\pm$    0.03 \\ 
B225       &   11.38 $\pm$    1.74 &   -0.44 $\pm$    0.03 \\
B234       &   10.87 $\pm$    1.10 &   -0.73 $\pm$    0.02 \\ 
B292       &    4.09 $\pm$    0.47 &   -1.54 $\pm$    0.04 \\
B301       &   15.89 $\pm$    4.97 &   -1.19 $\pm$    0.12 \\
B302       &   11.23 $\pm$    1.12 &   -1.49 $\pm$    0.05 \\ 
B304       &    8.53 $\pm$    2.22 &   -1.27 $\pm$    0.06 \\
B305       &    0.98 $\pm$    0.10 &    0.07 $\pm$    0.02 \\ 
B307       &    1.68 $\pm$    0.17 &    0.02 $\pm$    0.02 \\ 
B310       &    5.56 $\pm$    0.66 &   -1.49 $\pm$    0.05 \\
B313       &   12.67 $\pm$    1.27 &   -0.83 $\pm$    0.02 \\ 
B314       &    0.79 $\pm$    0.08 &   -0.12 $\pm$    0.02 \\ 
B316       &    1.46 $\pm$    0.26 &   -0.00 $\pm$    0.13 \\
B321       &    0.20 $\pm$    0.02 &   -0.07 $\pm$    0.06 \\ 
B322       &    0.14 $\pm$    0.09 &   -0.32 $\pm$    0.31 \\ 
B324       &    1.00 $\pm$    0.10 &   -0.13 $\pm$    0.09 \\
B327       &    0.90 $\pm$    0.54 &   -1.56 $\pm$    0.09 \\
B328       &    2.58 $\pm$    0.81 &   -1.60 $\pm$    0.05 \\
B331       &    5.97 $\pm$    1.34 &   -0.64 $\pm$    0.07 \\
B337       &    1.91 $\pm$    0.98 &   -0.58 $\pm$    0.32 \\
B347       &    8.06 $\pm$    1.29 &   -2.05 $\pm$    0.04 \\
B350       &    8.70 $\pm$    1.39 &   -1.66 $\pm$    0.04 \\
B354       &   11.43 $\pm$    1.14 &   -2.01 $\pm$    0.03 \\ 
B365       &    9.01 $\pm$    1.73 &   -1.34 $\pm$    0.04 \\
B380       &    0.58 $\pm$    0.06 &   -0.06 $\pm$    0.03 \\ 
B383       &   13.97 $\pm$    1.37 &   -0.57 $\pm$    0.03 \\
B393       &   15.71 $\pm$    1.58 &   -1.00 $\pm$    0.02 \\ 
B398       &   16.30 $\pm$    1.78 &   -0.60 $\pm$    0.03 \\
B401       &    8.49 $\pm$    0.85 &   -2.22 $\pm$    0.04 \\ 
MGC1       &   16.37 $\pm$    2.93 &   -1.36 $\pm$    0.03 \\
MGC5      &    9.99 $\pm$    1.00 &   -1.17 $\pm$    0.01 \\ 
MGC10     &   11.60 $\pm$    1.16 &   -1.76 $\pm$    0.01 \\ 
NB16       &    2.90 $\pm$    0.56 &   -1.09 $\pm$    0.11 \\
NB89       &   12.14 $\pm$    2.42 &   -0.77 $\pm$    0.05 \\
\hline
\end{tabular}
\end{table}

\addtocounter{table}{1}

\begin{figure}[h]
\begin{center}
\includegraphics[width=1.0\columnwidth]{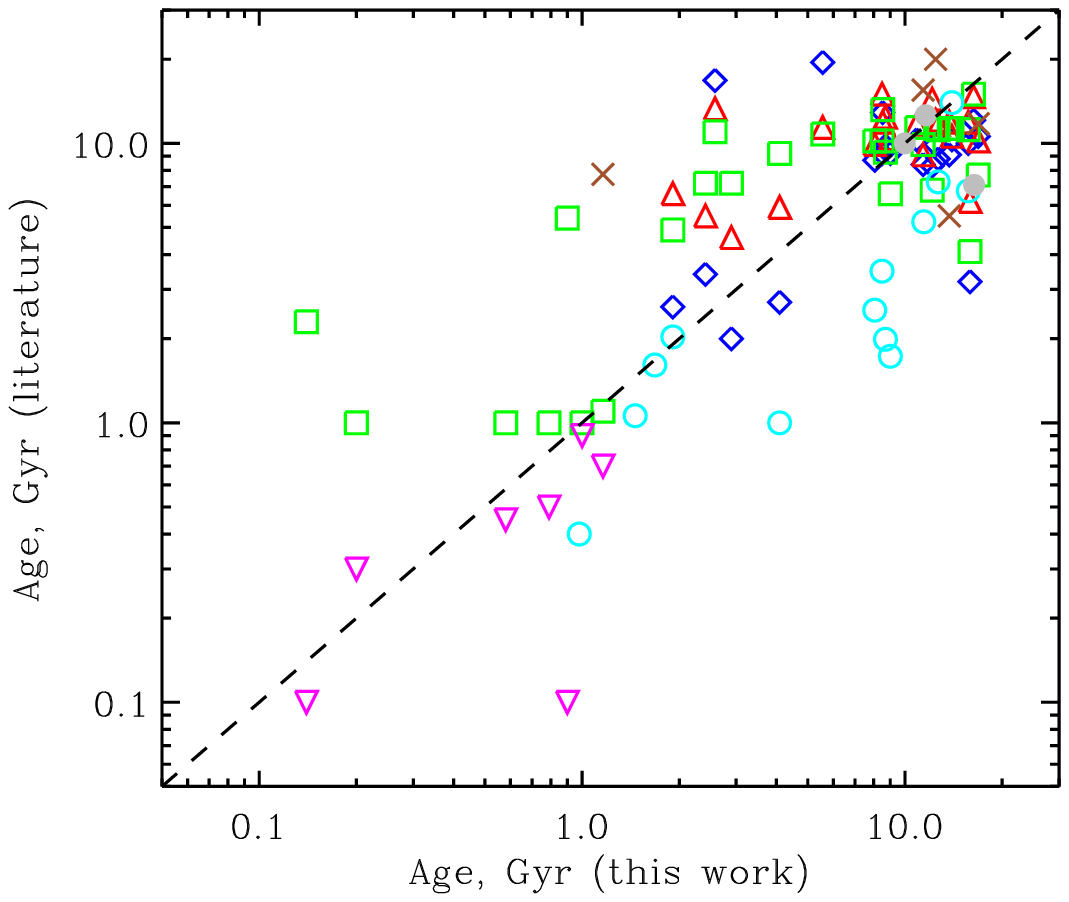}
\caption{\label{f:m31litage}Comparison between our derived ages and those published in literature for GCs in M31. 
Each coloured-symbol correspond to a different literature work, following Table \ref{tab_lit_m31}: 
(a) blue diamond correspond to results by \citet{beasley+05} (models by \citealt{BC03});
(b) red upward triangles to \citet{beasley+05} (models by \citealt{tmb03,tmk04});
(c) green squares to \citet{puzia+05};
(d) brown crosses to \citet{fan+06};
(e) magenta downward triangles to \citet{beasley+04};
(f) cyan open circles to \citet{wang+10}, and;
(g) grey filled circles to \citet{abrito+09}. As labeled in Table \ref{tab_lit_m31}, references (a-c) and (e) present the
results based on photometric measurements, while references (d), (f) and (g) above present the results based on spectral indices.
This plot is best seen in colour (online version).}
\end{center}
\end{figure}

Many of our GCs have been studied in literature, some by isochrones
fitting to CMDs, or via spectral indices. 
A non-exhaustive 
list of results from literature is presented in Table \ref{tab_lit_m31}. 
We compare our ages with those in the literature in Fig. \ref{f:m31litage}. 
From 
this figure, we conclude there is, on average, general agreement between our results and those 
in the literature, but with large dispersions. 

\begin{figure}[h]
\begin{center}
\includegraphics[width=1.0\columnwidth, clip=true]{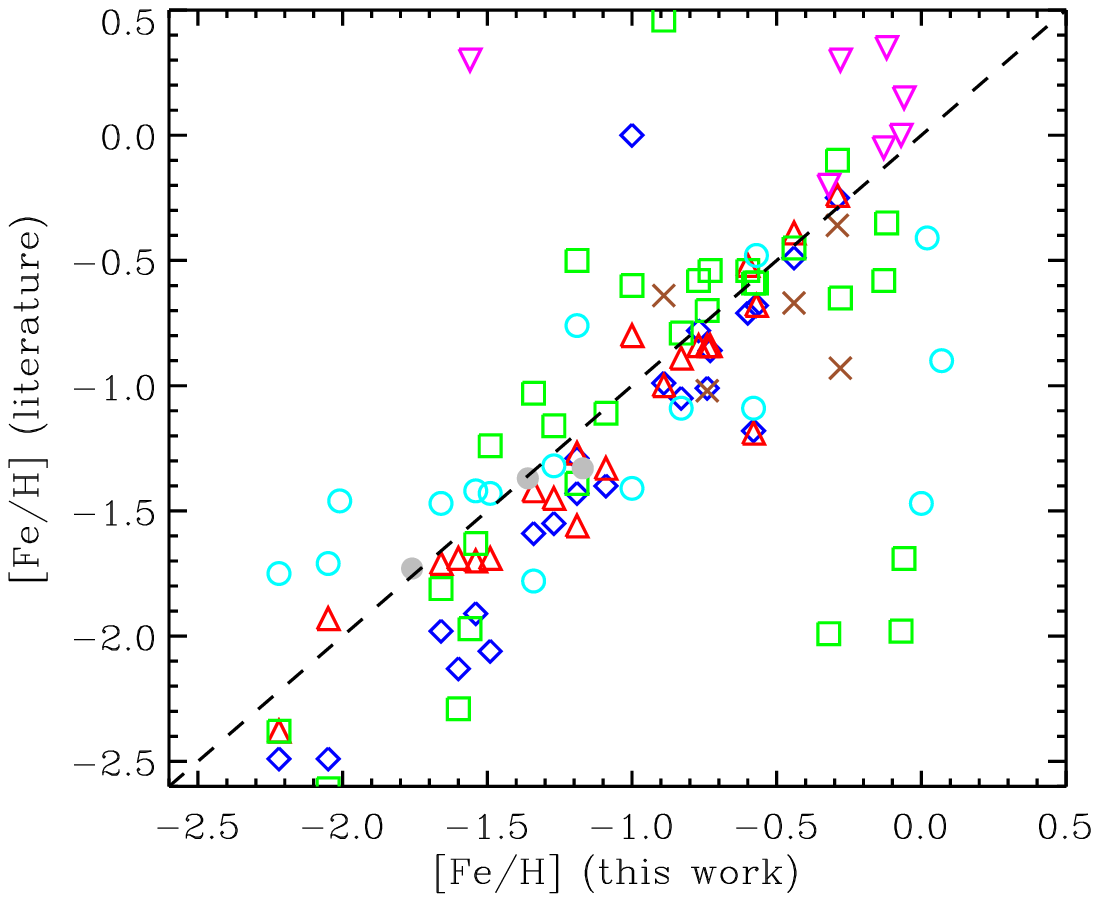}
\caption{\label{f:m31litfeh}Comparison between our derived metallicities [Fe/H] and those published in literature for GCs
in M31. The colour and symbol codes are the same as in Fig. \ref{f:m31litage}. This plot is best seen in colour (online version).}
\end{center}
\end{figure}

For the objects in common between this work and literature
 (see Table \ref{tab_lit_m31} and Fig
\ref{f:m31litfeh}'s caption for details), 
we find a mean difference of 2.45 Gyr with a r.m.s. of 5.39
Gyr for photometry-based ages; and 0.06 Gyr with a r.m.s. of 4.70 Gyr for spectral indices ages. The
agreement with metallicities (Fig. \ref{f:m31litfeh}) is tighter: 
 we determine a mean [Fe/H] difference
of 0.21 dex (photometry-based, with a r.m.s. of 0.54 dex) and of 0.10 dex with a
r.m.s. of 0.50 dex for those values determined through spectral indices.
We also compared the age and metallicity measurements for those M31 GCs in
common with the studies of \citet{beasley+05} and \citet{puzia+05}. Both groups
have used the Lick-index method but different SSP models. The mean age and
[Fe/H] difference between \citet{beasley+05} and \citet{puzia+05} results is 
of --0.60 Gyr (r.m.s. of 3.35 Gyr) and of --0.28 dex (r.m.s. =
0.48 dex), respectively.

We derive intermediate ages (2 -- 8 Gyr) for 7 GCs in our sample.
\citet{cald+11} recently analysed a large sample of M31 clusters and they found that
most of the intermediate age clusters (as analysed by previous work in literature) were 
actually old metal-poor clusters. Their analysis is based on spectral indices and models by
\citet{schiavon07}. 
Using spectral fitting and \citet{vazdekis+10} SSP models, 
our results agree better with previous work which obtain 
intermediate ages \citep[][]{beasley+04,beasley+05,puzia+05}\footnote{But see as well \citet{strader+09}, who performed mass-to-light ratio analysis and favored old instead of intermediate ages for some of these GCs..}.
 
We have three outer halo GCs in our sample that were
previously analysed in the study of \citet{abrito+09}. For MGC1, MGC5, and MGC 10
we find ages and metallicities of, respectively, (16.37, 9.99, and 11.60 Gyr)
and (--1.36, --1.17, and --1.76 dex). Except for MGC1, for which we have found a
higher age, there is a good agreement with the values of (7.10, 10.00, and 12.60
Gyr)
and (--1.37, --1.33, and --1.73 dex) reported by \citet{abrito+09}, respectively. 
MGC1 has also been recently investigated photometrically and
spectroscopically. Through the CMD analysis, \citet{mackey+10} conclude that MGC1
is old (12.5 to 12.7 Gyr) and metal-poor ([Fe/H] = --2.3 dex). Spectroscopically
through $\chi^{2}$-minimisation fittings, \citet {fan+12} also estimated an old age
(13.3 Gyr) for MGC1 but average metallicities ranging from [Fe/H] = --2.06 to
[Fe/H] = --1.76, depending on the SSP models used as well as on the stellar
evolutionary tracks employed. In addition, while \citet{fan+12} adopted a +0.42 dex enhancement of
$\alpha$ elements in their analysis, \citet{abrito+09} have used a solar mixture,
which could also account for the main
differences in [Fe/H] between the different works.

Photometric, spectral indices or spectral fitting methods 
differ not only in their 
technical details, but also their underlying stellar population 
models differ in their choice of evolutionary tracks, the libraries 
of stellar spectra and in their implementation details, such as 
interpolations. 
\citet{coelho+09}, for example, made an analysis of the stellar population in M32 and has shown that using the same method with 
different SSP models result in different age and metallicity distributions; they conclude that choosing different SSP
models from the literature might yield different results of age and metallicity at the quantitative level, even though a qualitative agreement is met. \citet{dias+10} also found evidence of this dependence of the results on the choice of SSP models when studying integrated spectra of
GCs in the Small Magellanic Cloud.  
As we tested our method with GCs in the Milky Way,
we believe that our techniques and models
are reliable, with the caveat that some objects ($\sim$12\% from the galactic sample) might have their ages underestimated by $\sim$~3~Gyr.

\citet{beasley+05,puzia+05} derive [$\alpha$/Fe] values as well as age and metallicities 
(in our Table \ref{tab_lit_m31} we converted their [Z/H] to [Fe/H] through equation 4 in \citealt{tmb03}).
Although there has been progress in measuring abundance patterns from spectral fitting 
\citep{walcher+09} we have not attempted it in this work as the models currently available do 
not cover the parameter space needed. We intend to investigate this in 
a future work, with a larger sample.

\subsection{Comparing the two GC systems}

\begin{figure}[h]
\begin{center}
\includegraphics[width=1.0\columnwidth]{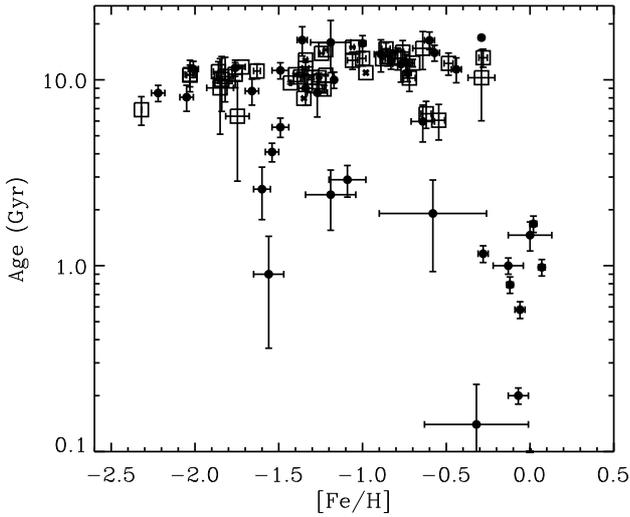}
\caption{\label{fig_agefe}Age-metallicity relation obtained for M31 (filled circles) and Galactic GCs (open
squares). Refer to the text of the paper for more details.}
\end{center}
\end{figure}

We show in Fig. \ref{fig_agefe} the age-[Fe/H] relation we obtain for our two
samples,
the Galactic GCs denoted by open squares and M31 GCs denoted by filled circles. 
It can be easily seen that the Galactic sample are consistent with old ages and 
a flat
age-[Fe/H] relation, as expected from earlier studies on the GC system of our
Galaxy.
Using a sample of 54 Galactic GCs with high-quality CMDs and ages obtained
photometrically, \citet{fraix-burnet+09} found an age-metallicity relation for the halo GCs. 
According to
their analysis the age-[Fe/H] relation goes from [Fe/H] = --2.0 dex 12 Gyr to
[Fe/H] = --1.3 dex, 9 Gyr ago. Nevertheless, their results are also consistent
with a fast increase in metallicity (--1.9 $\leq$ [Fe/H] $\leq$ --1.4), at
approximately 10--11 Gyr ago. More recently, \citet{forbes_bridges10} used new observations from the 
Hubble Space Telescope presented in \citet{marin-franch+09} to
compile a high-quality database of 93 Galactic GCs. With the larger sample, \citet{forbes_bridges10}
showed that, in fact, the Galactic GCs can be divided into two groups. The first group is
an
old population of GCs, compatible with a rapid formation scenario of the
Galactic halo, presenting a
flat age-[Fe/H] relation. The second group, however, presents some younger
objects displaying an age-[Fe/H]
relation which is associated with disrupted dwarf
galaxies. 

However, while the Galactic GCs are predominantly old, M31 shows not only young objects ($\leq$ 1 Gyr), but also a number of intermediate
age and old GCs \citep[][]{beasley+05,puzia+05}.
These findings are also confirmed in our analysis as shown in Table
\ref{tab_res_m31}. 
From Fig. \ref{fig_agefe} (filled symbols), we
see that, within the uncertainties, both old GCs in the Galaxy and in M31
present similar flat age-[Fe/H]
relations. 

By contrast, we see that M31's 
intermediate/young population (age $<$ 8 Gyr) follows a roughly linear age--[Fe/H] relation
in the sense that the young GCs are also more metal-rich.  These remarkable
differences in age-[Fe/H] between the Galactic and M31 GC systems suggest that 
while the latter has likely experienced a recent, active merger history, the
former has instead experienced a quiet star formation history over the last 10
Gyr \citep[but see][for a recent discussion on the topic]{forbes_bridges10}. We notice, however,
that a larger sample of young GCs in M31 has to be targeted to confirm (or not) the age-[Fe/H]
relation we recover in this work.

The relation between the GC system's stellar population and kinematic properties can shed light into formation of structures in a galaxy. For example, \citet{perrett+02} investigated the kinematic properties of several hundred GCs in M31. They concluded that the metal-rich GCs present a centrally concentrated spatial distribution with a high rotation amplitude, consistent with a bulge population. Whereas the metal-poor GCs tend to be less spatially concentrated and were also found to have a strong rotation signature (see as well discussions in \citealt{huchra+91,bekki10,lee+08,morrison+11} and references therein).

\begin{figure}[h]
\begin{center}
\includegraphics[width=1.0\columnwidth]{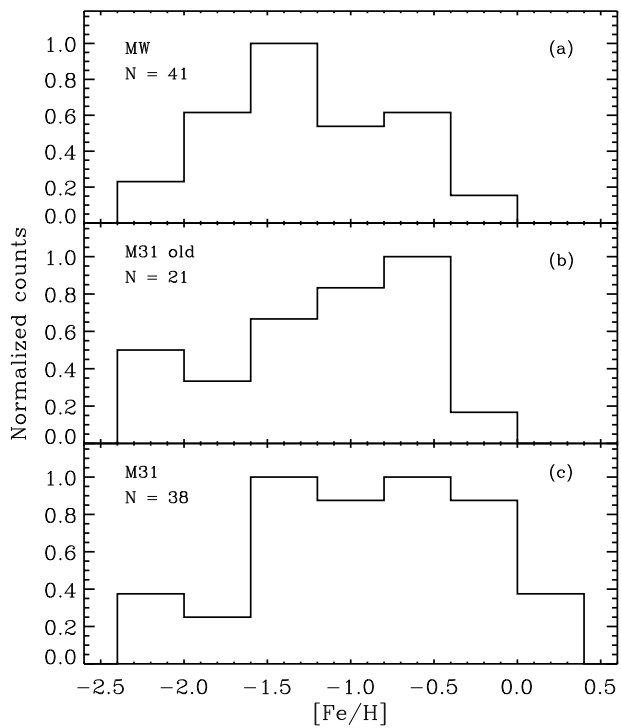}
\caption{\label{f:mdf} The metallicity distribution function for (a) Galactic GCs, (b) old (ages $\geq$ 8 Gyr) M31 GCs,
and (c) all (old+young) M31 GCs as analysed in this
work in
bins of 0.40 dex wide.}
\end{center}
\end{figure}

Also, the dis(similarities) between the metallicity distribution function (MDF) of the
Galactic and M31 GC systems have been widely debated in literature. 
In Fig. \ref{f:mdf}, we present the MDF we recover from our own analysis for both systems.
To our knowledge, this is the first direct comparison of [Fe/H] distributions in both Galactic and M31 GC systems
employing the same methods and techniques, regardless of the GC's position in the
galaxy. In the top
panel, we see that the MDF obtained from our integrated metallicities for the Galactic GCs is clearly
bimodal, a result that goes back to the seminal paper by
\citet{zinn85}, who was the first to propose the existence of two sub-populations in
the Galactic GC system \citep[see also][for a recent discussion]{bica+06}. 
The KMM mixture modelling algorithm \citep{ashman+94} suggests 
statistically convincing evidence of bimodality in the Galactic GC system (at
better than the 99\% confidence level).

For M31, however, the shape and distribution of the MDF is still
controversial. 
While some
authors propose that the MDF of M31 GCs presents two sub-populations -- one with a metallicity peak at
[Fe/H] = --1.57 that is associated to the galaxy's halo, and other one peaking at [Fe/H] =
--0.61, which is structurally associated to the galaxy's bulge \citep[see, e.g.,][]{ashman+93,barmby+00,fan+08} -- other
authors
suggest that the bimodality is not present at all or it is weakly detected \citep[e.g.][]{cald+11}.

From our own analysis for both Galactic and M31 GCs using the same methods and
techniques, we obtain that both systems, regardless the age
differences, cover approximately the same [Fe/H] range. Within our limited
sample, the old M31 GCs extend from [Fe/H] = --2.2 up to [Fe/H] = --0.3, while
the younger population goes from [Fe/H] = --2.2 up to 0.1 
dex\footnote{Note that the lowest metallicity in the SSP models grid is [Fe/H] = --2.3. Therefore, if there is tail in the MDF extending to lower metallicities, we would not be able to see it in this analysis.}. 
Furthermore, the KMM algorithm applied to the old
GCs in M31 does not support bimodality (the probability is less than 77\%), with the
number of metal-poor and metal-rich objects being almost the same in the galaxy.
We note, however, that M31's GC system is more than a factor of 2 larger than the
Milky Way's. In addition, our sample is biased not only by the number of objects studied but
also by their position in the galaxy. While our sample is biased towards
disk/bulge objects, more GCs in the halo of M31 need to be targeted to better
understand M31's MDF shape and properly compare it with that of the Galaxy.   

\section{Summary and Conclusions}
Spectroscopic ages and metallicities were derived for a sample of 38 GCs in M31, drawn from the observations of \citet{beasley+04} and \citet{abrito+09}.
These parameters were obtained by fitting the observed integrated spectra
to SSP models by \citet{vazdekis+10} using the spectral fitting code \emph{ULySS} \citep{ulyss}.
To our knowledge, this is the first time that full spectrum fitting is used in deriving stellar population parameters in 
M31 GCs.

We tested the reliability of our analysis by fitting the integrated spectra of Galactic GCs presented in \citet{schiavon+05}. In six cases,
out of 34 objects for which we obtained CMD ages from the literature, the spectroscopic ages do not match
the ages drawn from CMD analysis. 
In the case of NGC~5286 (and possibly NGC~6752), we suspect that
contamination from foreground stars hampered the observations. This is unlikely
to be an issue for extra-galactic clusters. NGC~2808 is a striking case of a
cluster with triple main sequences and complex HB morphology, and would deserve
a more detailed modelling than SSP fitting. For the remaining four clusters, ages are underestimated by $\sim$2--3 Gyr and we cannot provide a robust explanation for these differences.
We did not find evidence of a correlation with either contamination from CNONa abundance variations or a specific HB morphology.

The spectroscopic integrated metallicities derived with spectral fitting were compared to the compilations by \citet{schiavon+05} and \citet{carretta+09}. Our results agree well with the higher resolution stellar analysis.
 
As for M31, we obtain a large range of ages (from $\sim$150~Myr to the age of the
Universe) and metallicities (--2.2 $\leq$ [Fe/H] $\leq$ +0.1). We confirm previous results in the literature that find young globular clusters in M31, in contrast with
the globular cluster system in our own Milky Way.
We find an age-metallicity relation that can be described as having two components:
an old population with a flat age-[Fe/H] relation, possibly associated with the halo and/or bulge in analogy to what is seen in
the 
Milky Way, and a second one with a roughly linear relation between age and metallicity (higher metallicities corresponding to younger ages).

Finally, our analyses do not support [Fe/H] bimodality in the M31 globular cluster system, but further investigation with a larger and unbiased sample is necessary to
better interpret the star formation history of M31 when compared to our own Galaxy.


\begin{acknowledgements}
EC and PC are thankful to Mina Koleva for her constant help on using
the {\it ULySS} package. PC acknowledges financial support from FAPESP project
2008/58406-4. AB acknowledges support from Australian Research Council (Super Science Fellowship, FS110200016). 
JB acknowledges support from NSF grants AST-1211995 and AST-1109878.
\end{acknowledgements}


\bibliographystyle{aa}
\bibliography{bibliography}

\begin{thebibliography}{81}
\expandafter\ifx\csname natexlab\endcsname\relax\def\natexlab#1{#1}\fi

\bibitem[{{Alves-Brito} {et~al.}(2009){Alves-Brito}, {Forbes}, {Mendel}, {Hau},
  \& {Murphy}}]{abrito+09}
{Alves-Brito}, A., {Forbes}, D.~A., {Mendel}, J.~T., {Hau}, G.~K.~T., \&
  {Murphy}, M.~T. 2009, \mnras, 395, L34

\bibitem[{{Alves-Brito} {et~al.}(2011){Alves-Brito}, {Hau}, {Forbes},
  {Spitler}, {Strader}, {Brodie}, \& {Rhode}}]{abrito+11}
{Alves-Brito}, A., {Hau}, G.~K.~T., {Forbes}, D.~A., {et~al.} 2011, \mnras,
  417, 1823

\bibitem[{{Ashman} \& {Bird}(1993)}]{ashman+93}
{Ashman}, K.~M. \& {Bird}, C.~M. 1993, \aj, 106, 2281

\bibitem[{{Ashman} {et~al.}(1994){Ashman}, {Bird}, \& {Zepf}}]{ashman+94}
{Ashman}, K.~M., {Bird}, C.~M., \& {Zepf}, S.~E. 1994, \aj, 108, 2348

\bibitem[{{Barmby} {et~al.}(2000){Barmby}, {Huchra}, {Brodie}, {Forbes},
  {Schroder}, \& {Grillmair}}]{barmby+00}
{Barmby}, P., {Huchra}, J.~P., {Brodie}, J.~P., {et~al.} 2000, \aj, 119, 727

\bibitem[{{Beasley} {et~al.}(2004){Beasley}, {Brodie}, {Strader}, {Forbes},
  {Proctor}, {Barmby}, \& {Huchra}}]{beasley+04}
{Beasley}, M.~A., {Brodie}, J.~P., {Strader}, J., {et~al.} 2004, \aj, 128, 1623

\bibitem[{{Beasley} {et~al.}(2005){Beasley}, {Brodie}, {Strader}, {Forbes},
  {Proctor}, {Barmby}, \& {Huchra}}]{beasley+05}
{Beasley}, M.~A., {Brodie}, J.~P., {Strader}, J., {et~al.} 2005, \aj, 129, 1412

\bibitem[{{Beaulieu} {et~al.}(2001){Beaulieu}, {Gilmore}, {Elson}, {Johnson},
  {Santiago}, {Sigurdsson}, \& {Tanvir}}]{beaulieu+01}
{Beaulieu}, S.~F., {Gilmore}, G., {Elson}, R.~A.~W., {et~al.} 2001, \aj, 121,
  2618

\bibitem[{{Bekki}(2010)}]{bekki10}
{Bekki}, K. 2010, \mnras, 401, L58

\bibitem[{{Bica} {et~al.}(2006){Bica}, {Bonatto}, {Barbuy}, \&
  {Ortolani}}]{bica+06}
{Bica}, E., {Bonatto}, C., {Barbuy}, B., \& {Ortolani}, S. 2006, \aap, 450, 105

\bibitem[{{Brodie} \& {Huchra}(1990)}]{brodie_huchra90}
{Brodie}, J.~P. \& {Huchra}, J.~P. 1990, \apj, 362, 503

\bibitem[{{Bruzual} \& {Charlot}(2003)}]{BC03}
{Bruzual}, G. \& {Charlot}, S. 2003, MNRAS, 344, 1000

\bibitem[{{Buonanno} {et~al.}(1998){Buonanno}, {Corsi}, {Pulone}, {Fusi Pecci},
  \& {Bellazzini}}]{buonanno+98}
{Buonanno}, R., {Corsi}, C.~E., {Pulone}, L., {Fusi Pecci}, F., \&
  {Bellazzini}, M. 1998, \aap, 333, 505

\bibitem[{{Caldwell} {et~al.}(2011){Caldwell}, {Schiavon}, {Morrison}, {Rose},
  \& {Harding}}]{cald+11}
{Caldwell}, N., {Schiavon}, R., {Morrison}, H., {Rose}, J.~A., \& {Harding}, P.
  2011, \aj, 141, 61

\bibitem[{{Carretta} {et~al.}(2009){Carretta}, {Bragaglia}, {Gratton},
  {D'Orazi}, \& {Lucatello}}]{carretta+09}
{Carretta}, E., {Bragaglia}, A., {Gratton}, R., {D'Orazi}, V., \& {Lucatello},
  S. 2009, \aap, 508, 695

\bibitem[{{Carretta} \& {Gratton}(1997)}]{carreta_gratton97}
{Carretta}, E. \& {Gratton}, R.~G. 1997, \aaps, 121, 95

\bibitem[{{Cenarro} {et~al.}(2007){Cenarro}, {Peletier},
  {S{\'a}nchez-Bl{\'a}zquez}, {Selam}, {Toloba}, {Cardiel},
  {Falc{\'o}n-Barroso}, {Gorgas}, {Jim{\'e}nez-Vicente}, \&
  {Vazdekis}}]{MILES2}
{Cenarro}, A.~J., {Peletier}, R.~F., {S{\'a}nchez-Bl{\'a}zquez}, P., {et~al.}
  2007, \mnras, 374, 664

\bibitem[{{Chaboyer} {et~al.}(1998){Chaboyer}, {Demarque}, {Kernan}, \&
  {Krauss}}]{chaboyer+98}
{Chaboyer}, B., {Demarque}, P., {Kernan}, P.~J., \& {Krauss}, L.~M. 1998, \apj,
  494, 96

\bibitem[{{Cid Fernandes} \& {Gonz{\'a}lez Delgado}(2010)}]{cid_delgado10}
{Cid Fernandes}, R. \& {Gonz{\'a}lez Delgado}, R.~M. 2010, \mnras, 403, 780

\bibitem[{{Coelho} {et~al.}(2007){Coelho}, {Bruzual}, {Charlot}, {Weiss},
  {Barbuy}, \& {Ferguson}}]{coelho+07}
{Coelho}, P., {Bruzual}, G., {Charlot}, S., {et~al.} 2007, \mnras, 382, 498

\bibitem[{{Coelho} {et~al.}(2009){Coelho}, {Mendes de Oliveira}, \& {Cid
  Fernandes}}]{coelho+09}
{Coelho}, P., {Mendes de Oliveira}, C., \& {Cid Fernandes}, R. 2009, \mnras,
  396, 624

\bibitem[{{Coelho} {et~al.}(2012){Coelho}, {Percival}, \&
  {Salaris}}]{coelho+12proc}
{Coelho}, P., {Percival}, S., \& {Salaris}, M. 2012, ArXiv e-prints

\bibitem[{{Coelho} {et~al.}(2011){Coelho}, {Percival}, \&
  {Salaris}}]{coelho+11}
{Coelho}, P., {Percival}, S.~M., \& {Salaris}, M. 2011, ApJ, 734, 72

\bibitem[{{Conroy} \& {Gunn}(2010)}]{conroy_gunn10}
{Conroy}, C. \& {Gunn}, J.~E. 2010, \apj, 712, 833

\bibitem[{{De Angeli} {et~al.}(2005){De Angeli}, {Piotto}, {Cassisi}, {Busso},
  {Recio-Blanco}, {Salaris}, {Aparicio}, \& {Rosenberg}}]{deangeli+05}
{De Angeli}, F., {Piotto}, G., {Cassisi}, S., {et~al.} 2005, \aj, 130, 116

\bibitem[{{de Freitas Pacheco} \& {Barbuy}(1995)}]{pacheco_barbuy95}
{de Freitas Pacheco}, J.~A. \& {Barbuy}, B. 1995, \aap, 302, 718

\bibitem[{{Delgado} {et~al.}(2005){Delgado}, {Cervi{\~n}o}, {Martins},
  {Leitherer}, \& {Hauschildt}}]{delgado+05}
{Delgado}, R.~M.~G., {Cervi{\~n}o}, M., {Martins}, L.~P., {Leitherer}, C., \&
  {Hauschildt}, P.~H. 2005, MNRAS, 357, 945

\bibitem[{{Dias} {et~al.}(2010){Dias}, {Coelho}, {Barbuy}, {Kerber}, \&
  {Idiart}}]{dias+10}
{Dias}, B., {Coelho}, P., {Barbuy}, B., {Kerber}, L., \& {Idiart}, T. 2010,
  \aap, 520, A85+

\bibitem[{{Fan} \& {de Grijs}(2012)}]{fan+12}
{Fan}, Z. \& {de Grijs}, R. 2012, \mnras, 424, 2009

\bibitem[{{Fan} {et~al.}(2006){Fan}, {Ma}, {de Grijs}, {Yang}, \&
  {Zhou}}]{fan+06}
{Fan}, Z., {Ma}, J., {de Grijs}, R., {Yang}, Y., \& {Zhou}, X. 2006, \mnras,
  371, 1648

\bibitem[{{Fan} {et~al.}(2008){Fan}, {Ma}, {de Grijs}, \& {Zhou}}]{fan+08}
{Fan}, Z., {Ma}, J., {de Grijs}, R., \& {Zhou}, X. 2008, \mnras, 385, 1973

\bibitem[{{Forbes} \& {Bridges}(2010)}]{forbes_bridges10}
{Forbes}, D.~A. \& {Bridges}, T. 2010, \mnras, 404, 1203

\bibitem[{{Fraix-Burnet} {et~al.}(2009){Fraix-Burnet}, {Davoust}, \&
  {Charbonnel}}]{fraix-burnet+09}
{Fraix-Burnet}, D., {Davoust}, E., \& {Charbonnel}, C. 2009, \mnras, 398, 1706

\bibitem[{{Gallart} {et~al.}(2005){Gallart}, {Zoccali}, \&
  {Aparicio}}]{gallart+05}
{Gallart}, C., {Zoccali}, M., \& {Aparicio}, A. 2005, \araa, 43, 387

\bibitem[{{Galleti} {et~al.}(2004){Galleti}, {Federici}, {Bellazzini}, {Fusi
  Pecci}, \& {Macrina}}]{galleti+04}
{Galleti}, S., {Federici}, L., {Bellazzini}, M., {Fusi Pecci}, F., \&
  {Macrina}, S. 2004, \aap, 416, 917

\bibitem[{{Girardi} {et~al.}(2000){Girardi}, {Bressan}, {Bertelli}, \&
  {Chiosi}}]{girardi+00}
{Girardi}, L., {Bressan}, A., {Bertelli}, G., \& {Chiosi}, C. 2000, A\&AS, 141,
  371

\bibitem[{{Gratton} {et~al.}(2012){Gratton}, {Carretta}, \&
  {Bragaglia}}]{gratton+12}
{Gratton}, R.~G., {Carretta}, E., \& {Bragaglia}, A. 2012, \aapr, 20, 50

\bibitem[{{Gratton} {et~al.}(2010){Gratton}, {Carretta}, {Bragaglia},
  {Lucatello}, \& {D'Orazi}}]{gratton+10a}
{Gratton}, R.~G., {Carretta}, E., {Bragaglia}, A., {Lucatello}, S., \&
  {D'Orazi}, V. 2010, \aap, 517, A81

\bibitem[{{Holland}(1998)}]{holland98}
{Holland}, S. 1998, \aj, 115, 1916

\bibitem[{{Huchra} {et~al.}(1991){Huchra}, {Brodie}, \& {Kent}}]{huchra+91}
{Huchra}, J.~P., {Brodie}, J.~P., \& {Kent}, S.~M. 1991, \apj, 370, 495

\bibitem[{{Jarosik} {et~al.}(2011){Jarosik}, {Bennett}, {Dunkley}, {Gold},
  {Greason}, {Halpern}, {Hill}, {Hinshaw}, {Kogut}, {Komatsu}, {Larson},
  {Limon}, {Meyer}, {Nolta}, {Odegard}, {Page}, {Smith}, {Spergel}, {Tucker},
  {Weiland}, {Wollack}, \& {Wright}}]{jarosik+11}
{Jarosik}, N., {Bennett}, C.~L., {Dunkley}, J., {et~al.} 2011, \apjs, 192, 14

\bibitem[{{Kent} {et~al.}(1989){Kent}, {Huchra}, \& {Stauffer}}]{kent+89}
{Kent}, S.~M., {Huchra}, J.~P., \& {Stauffer}, J. 1989, \aj, 98, 2080

\bibitem[{{Koleva} {et~al.}(2009){Koleva}, {Prugniel}, {Bouchard}, \&
  {Wu}}]{ulyss}
{Koleva}, M., {Prugniel}, P., {Bouchard}, A., \& {Wu}, Y. 2009, \aap, 501, 1269

\bibitem[{{Koleva} {et~al.}(2008){Koleva}, {Prugniel}, {Ocvirk}, {Le Borgne},
  \& {Soubiran}}]{koleva+08}
{Koleva}, M., {Prugniel}, P., {Ocvirk}, P., {Le Borgne}, D., \& {Soubiran}, C.
  2008, \mnras, 385, 1998

\bibitem[{{Kraft} \& {Ivans}(2003)}]{kraft_ivans03}
{Kraft}, R.~P. \& {Ivans}, I.~I. 2003, \pasp, 115, 143

\bibitem[{{Le Borgne} {et~al.}(2004){Le Borgne}, {Rocca-Volmerange},
  {Prugniel}, {Lan{\c c}on}, {Fioc}, \& {Soubiran}}]{PEGASE-HR}
{Le Borgne}, D., {Rocca-Volmerange}, B., {Prugniel}, P., {et~al.} 2004, A\&A,
  425, 881

\bibitem[{{Lee} {et~al.}(2009){Lee}, {Worthey}, {Dotter}, {Chaboyer},
  {Jevremovi{\'c}}, {Baron}, {Briley}, {Ferguson}, {Coelho}, \&
  {Trager}}]{lee+09}
{Lee}, H.-c., {Worthey}, G., {Dotter}, A., {et~al.} 2009, \apj, 694, 902

\bibitem[{{Lee} {et~al.}(2000){Lee}, {Yoon}, \& {Lee}}]{lee+00}
{Lee}, H.-c., {Yoon}, S.-J., \& {Lee}, Y.-W. 2000, \aj, 120, 998

\bibitem[{{Lee} {et~al.}(2008){Lee}, {Hwang}, {Kim}, {Park}, {Geisler},
  {Sarajedini}, \& {Harris}}]{lee+08}
{Lee}, M.~G., {Hwang}, H.~S., {Kim}, S.~C., {et~al.} 2008, \apj, 674, 886

\bibitem[{{Ma} {et~al.}(2007){Ma}, {Yang}, {Burstein}, {Fan}, {Wu}, {Zhou},
  {Wu}, {Jiang}, \& {Chen}}]{ma+07}
{Ma}, J., {Yang}, Y., {Burstein}, D., {et~al.} 2007, \apj, 659, 359

\bibitem[{{Mackey} {et~al.}(2010){Mackey}, {Huxor}, {Ferguson}, {Irwin},
  {Tanvir}, {McConnachie}, {Ibata}, {Chapman}, \& {Lewis}}]{mackey+10}
{Mackey}, A.~D., {Huxor}, A.~P., {Ferguson}, A.~M.~N., {et~al.} 2010, \apjl,
  717, L11

\bibitem[{{Maraston}(2005)}]{maraston05}
{Maraston}, C. 2005, \mnras, 362, 799

\bibitem[{{Mar{\'{\i}}n-Franch} {et~al.}(2009){Mar{\'{\i}}n-Franch},
  {Aparicio}, {Piotto}, {Rosenberg}, {Chaboyer}, {Sarajedini}, {Siegel},
  {Anderson}, {Bedin}, {Dotter}, {Hempel}, {King}, {Majewski}, {Milone},
  {Paust}, \& {Reid}}]{marin-franch+09}
{Mar{\'{\i}}n-Franch}, A., {Aparicio}, A., {Piotto}, G., {et~al.} 2009, \apj,
  694, 1498

\bibitem[{{Meissner} \& {Weiss}(2006)}]{meissner_weiss06}
{Meissner}, F. \& {Weiss}, A. 2006, \aap, 456, 1085

\bibitem[{{Mendel} {et~al.}(2007){Mendel}, {Proctor}, \& {Forbes}}]{mendel+07}
{Mendel}, J.~T., {Proctor}, R.~N., \& {Forbes}, D.~A. 2007, \mnras, 379, 1618

\bibitem[{{Milone} {et~al.}(2011){Milone}, {Sansom}, \&
  {S{\'a}nchez-Bl{\'a}zquez}}]{milone+11}
{Milone}, A.~D.~C., {Sansom}, A.~E., \& {S{\'a}nchez-Bl{\'a}zquez}, P. 2011,
  \mnras, 414, 1227

\bibitem[{{Momany} {et~al.}(2003){Momany}, {Ortolani}, {Held}, {Barbuy},
  {Bica}, {Renzini}, {Bedin}, {Rich}, \& {Marconi}}]{momany+03}
{Momany}, Y., {Ortolani}, S., {Held}, E.~V., {et~al.} 2003, \aap, 402, 607

\bibitem[{{Morrison} {et~al.}(2011){Morrison}, {Caldwell}, {Schiavon},
  {Athanassoula}, {Romanowsky}, \& {Harding}}]{morrison+11}
{Morrison}, H., {Caldwell}, N., {Schiavon}, R.~P., {et~al.} 2011, \apjl, 726,
  L9

\bibitem[{{Ocvirk}(2010)}]{ocvirk10}
{Ocvirk}, P. 2010, \apj, 709, 88

\bibitem[{{Percival} {et~al.}(2009){Percival}, {Salaris}, {Cassisi}, \&
  {Pietrinferni}}]{percival+09}
{Percival}, S.~M., {Salaris}, M., {Cassisi}, S., \& {Pietrinferni}, A. 2009,
  \apj, 690, 427

\bibitem[{{Perrett} {et~al.}(2002){Perrett}, {Bridges}, {Hanes}, {Irwin},
  {Brodie}, {Carter}, {Huchra}, \& {Watson}}]{perrett+02}
{Perrett}, K.~M., {Bridges}, T.~J., {Hanes}, D.~A., {et~al.} 2002, \aj, 123,
  2490

\bibitem[{{Piotto} {et~al.}(2007){Piotto}, {Bedin}, {Anderson}, {King},
  {Cassisi}, {Milone}, {Villanova}, {Pietrinferni}, \& {Renzini}}]{piotto+07}
{Piotto}, G., {Bedin}, L.~R., {Anderson}, J., {et~al.} 2007, \apjl, 661, L53

\bibitem[{{Proctor} {et~al.}(2004){Proctor}, {Forbes}, \&
  {Beasley}}]{proctor+04}
{Proctor}, R.~N., {Forbes}, D.~A., \& {Beasley}, M.~A. 2004, MNRAS, 355, 1327

\bibitem[{{Puzia} {et~al.}(2005){Puzia}, {Perrett}, \& {Bridges}}]{puzia+05}
{Puzia}, T.~H., {Perrett}, K.~M., \& {Bridges}, T.~J. 2005, \aap, 434, 909

\bibitem[{{Rutledge} {et~al.}(1997){Rutledge}, {Hesser}, {Stetson}, {Mateo},
  {Simard}, {Bolte}, {Friel}, \& {Copin}}]{rutledge+97}
{Rutledge}, G.~A., {Hesser}, J.~E., {Stetson}, P.~B., {et~al.} 1997, \pasp,
  109, 883

\bibitem[{{S{\'a}nchez-Bl{\'a}zquez} {et~al.}(2006){S{\'a}nchez-Bl{\'a}zquez},
  {Peletier}, {Jim{\'e}nez-Vicente}, {Cardiel}, {Cenarro},
  {Falc{\'o}n-Barroso}, {Gorgas}, {Selam}, \& {Vazdekis}}]{MILES1}
{S{\'a}nchez-Bl{\'a}zquez}, P., {Peletier}, R.~F., {Jim{\'e}nez-Vicente}, J.,
  {et~al.} 2006, \mnras, 371, 703

\bibitem[{{Schiavon}(2007)}]{schiavon07}
{Schiavon}, R.~P. 2007, \apjs, 171, 146

\bibitem[{{Schiavon} {et~al.}(2004){Schiavon}, {Rose}, {Courteau}, \&
  {MacArthur}}]{schiavon+04b}
{Schiavon}, R.~P., {Rose}, J.~A., {Courteau}, S., \& {MacArthur}, L.~A. 2004,
  \apjl, 608, L33

\bibitem[{Schiavon {et~al.}(2005)Schiavon, Rose, Courteau, \&
  MacArthur}]{schiavon+05}
Schiavon, R.~P., Rose, J.~A., Courteau, S., \& MacArthur, L.~A. 2005, ApJS,
  169, 163

\bibitem[{{Strader} {et~al.}(2009){Strader}, {Smith}, {Larsen}, {Brodie}, \&
  {Huchra}}]{strader+09}
{Strader}, J., {Smith}, G.~H., {Larsen}, S., {Brodie}, J.~P., \& {Huchra},
  J.~P. 2009, \aj, 138, 547

\bibitem[{{Thomas} {et~al.}(2003){Thomas}, {Maraston}, \& {Bender}}]{tmb03}
{Thomas}, D., {Maraston}, C., \& {Bender}, R. 2003, MNRAS, 339, 897

\bibitem[{{Thomas} {et~al.}(2004){Thomas}, {Maraston}, \& {Korn}}]{tmk04}
{Thomas}, D., {Maraston}, C., \& {Korn}, A. 2004, MNRAS, 351, L19

\bibitem[{{Usher} {et~al.}(2012){Usher}, {Forbes}, {Brodie}, {Foster},
  {Spitler}, {Arnold}, {Romanowsky}, {Strader}, \& {Pota}}]{usher+12}
{Usher}, C., {Forbes}, D.~A., {Brodie}, J.~P., {et~al.} 2012, \mnras, 426, 1475

\bibitem[{{Vazdekis} {et~al.}(2010){Vazdekis}, {S{\'a}nchez-Bl{\'a}zquez},
  {Falc{\'o}n-Barroso}, {Cenarro}, {Beasley}, {Cardiel}, {Gorgas}, \&
  {Peletier}}]{vazdekis+10}
{Vazdekis}, A., {S{\'a}nchez-Bl{\'a}zquez}, P., {Falc{\'o}n-Barroso}, J.,
  {et~al.} 2010, \mnras, 404, 1639

\bibitem[{{Walcher} {et~al.}(2009){Walcher}, {Coelho}, {Gallazzi}, \&
  {Charlot}}]{walcher+09}
{Walcher}, C.~J., {Coelho}, P., {Gallazzi}, A., \& {Charlot}, S. 2009, \mnras,
  398, L44

\bibitem[{{Wang} {et~al.}(2010){Wang}, {Fan}, {Ma}, {de Grijs}, \&
  {Zhou}}]{wang+10}
{Wang}, S., {Fan}, Z., {Ma}, J., {de Grijs}, R., \& {Zhou}, X. 2010, \aj, 139,
  1438

\bibitem[{{Wolf} {et~al.}(2007){Wolf}, {Drory}, {Gebhardt}, \&
  {Hill}}]{wolf+07}
{Wolf}, M.~J., {Drory}, N., {Gebhardt}, K., \& {Hill}, G.~J. 2007, \apj, 655,
  179

\bibitem[{Worthey {et~al.}(1994)Worthey, Faber, Gonzalez, \&
  Burstein}]{worthey+94}
Worthey, G., Faber, S.~M., Gonzalez, J.~J., \& Burstein, D. 1994, ApJS, 94, 687

\bibitem[{{Zinn}(1985)}]{zinn85}
{Zinn}, R. 1985, \apj, 293, 424

\bibitem[{{Zinn} \& {West}(1984)}]{zinn_west84}
{Zinn}, R. \& {West}, M.~J. 1984, \apjs, 55, 45

\bibitem[{{Zoccali} {et~al.}(2001){Zoccali}, {Renzini}, {Ortolani},
  {Bragaglia}, {Bohlin}, {Carretta}, {Ferraro}, {Gilmozzi}, {Holberg},
  {Marconi}, {Rich}, \& {Wesemael}}]{zoccali+01}
{Zoccali}, M., {Renzini}, A., {Ortolani}, S., {et~al.} 2001, \apj, 553, 733

\end{thebibliography}


\longtab{2}{
\begin{longtable}{ l c c c c c c}
\caption[Ages and metallicities for Galactic GCs]{\label{tab_res_mw} Ages and metallicities for Galactic GCs.}\\
\hline
   & \multicolumn{3}{c}{Literature}             &  & \multicolumn{2}{c}{This work}   \\
\cline{2-4} \cline{6-7} 
Cluster &  Age   & [Fe/H]$_C$ & [Fe/H]$_S$ & & Age   & [Fe/H]  \\
        & (Gyr)  &  (dex)     &  (dex)     & & (Gyr) & (dex)  \\
\hline
\endfirsthead
\caption{continued.}\\
\hline
   & \multicolumn{3}{c}{Literature}             &  & \multicolumn{2}{c}{This work}   \\
\cline{2-4} \cline{6-7} 
Cluster &  Age   & [Fe/H]$_C$ & [Fe/H]$_S$ & & Age   & [Fe/H]  \\
        & (Gyr)  &  (dex)     &  (dex)     & & (Gyr) & (dex)  \\
\hline
\endhead
\multicolumn{7}{r}{{Continued on next page}} \\ \hline
\endfoot
\hline
\endlastfoot
NGC~104  &  13.0 $\pm$ 2.6 $^a$ &   -0.76 $\pm$ 0.02  &   -0.70   &  &  13.09 $\pm$    1.76 &   -0.79 $\pm$    0.04 \\
NGC~1851 &  10.6 $\pm$ 2.1 $^a$ &   -1.18 $\pm$ 0.08  &   -1.21   &  &  10.26 $\pm$    0.16 &   -1.26 $\pm$    0.01 \\
NGC~1904 &  11.4 $\pm$ 2.4 $^a$ &   -1.58 $\pm$ 0.02  &   -1.55   &  &  11.58 $\pm$    0.22 &   -1.84 $\pm$    0.02 \\
        &                      &                     &           &  &  10.24 $\pm$    3.00 &   -1.88 $\pm$    0.05 \\
NGC~2298 &  12.6 $\pm$ 2.6 $^b$ &   -1.96 $\pm$ 0.04  &   -1.97   &  &  10.89 $\pm$    1.34 &   -2.00 $\pm$    0.03 \\
        &                      &                     &           &  &  10.32 $\pm$    1.54 &   -2.06 $\pm$    0.02 \\
NGC~2808 &  10.0 $\pm$ 2.2 $^a$ &   -1.18 $\pm$ 0.04  &   -1.29   &  &  13.90 $\pm$    0.07 &   -1.24 $\pm$    0.01 \\
        &                      &                     &           &  &  13.87 $\pm$    0.08 &   -1.25 $\pm$    0.01 \\
NGC~3201 &  13.0 $\pm$ 1.9 $^a$ &   -1.51 $\pm$ 0.02  &   -1.56   &  &   9.56 $\pm$    0.24 &   -1.35 $\pm$    0.01 \\
        &                      &                     &           &  &  11.75 $\pm$    0.11 &   -1.45 $\pm$    0.01 \\
NGC~5286 &  12.5 $\pm$ 2.5 $^b$ &   -1.70 $\pm$ 0.07  &   -1.51   &  &   6.04 $\pm$    4.46 &   -1.78 $\pm$    0.08 \\
        &                      &                     &           &  &   9.19 $\pm$    3.34 &   -1.79 $\pm$    0.07 \\
        &                      &                     &           &  &   4.01 $\pm$    2.97 &   -1.66 $\pm$    0.07 \\
NGC~5904 &  10.6 $\pm$ 2.2 $^a$ &   -1.33 $\pm$ 0.02  &   -1.26   &  &   9.26 $\pm$    0.11 &   -1.44 $\pm$    0.00 \\
        &                      &                     &           &  &  10.00 $\pm$    0.16 &   -1.41 $\pm$    0.01 \\
NGC~5927 &  12.1 $\pm$ 2.4 $^a$ &   -0.29 $\pm$ 0.07  &   -0.64   &  &  11.38 $\pm$    2.54 &   -0.48 $\pm$    0.04 \\
        &                      &                     &           &  &  12.87 $\pm$    1.73 &   -0.50 $\pm$    0.02 \\
        &                      &                     &           &  &  12.77 $\pm$    1.37 &   -0.48 $\pm$    0.02 \\
NGC~5946 &  11.7 $\pm$ 3.0 $^a$ &   -1.29 $\pm$ 0.14  &   -1.54   &  &  10.56 $\pm$    1.93 &   -1.76 $\pm$    0.04 \\
NGC~5986 &  11.6 $\pm$ 2.3 $^a$ &   -1.63 $\pm$ 0.08  &   -1.53   &  &  10.14 $\pm$    3.33 &   -1.84 $\pm$    0.05 \\
NGC~6121 &  11.9 $\pm$ 2.3 $^a$ &   -1.18 $\pm$ 0.02  &   -1.15   &  &   7.94 $\pm$    0.14 &   -1.35 $\pm$    0.01 \\
NGC~6171 &  13.0 $\pm$ 2.7 $^a$ &   -1.03 $\pm$ 0.02  &   -1.13   &  &  13.22 $\pm$    1.27 &   -1.07 $\pm$    0.03 \\
        &                      &                     &           &  &  12.23 $\pm$    1.47 &   -1.05 $\pm$    0.03 \\
NGC~6218 &  12.1 $\pm$ 2.5 $^a$ &   -1.33 $\pm$ 0.02  &   -1.32   &  &  11.08 $\pm$    0.94 &   -1.63 $\pm$    0.02 \\
NGC~6235 &  11.7 $\pm$ 3.0 $^a$ &   -1.38 $\pm$ 0.07  &   -1.36   &  &   9.81 $\pm$    1.07 &   -1.26 $\pm$    0.04 \\
NGC~6254 &  11.3 $\pm$ 2.3 $^a$ &   -1.57 $\pm$ 0.02  &   -1.51   &  &  11.73 $\pm$    0.05 &   -1.72 $\pm$    0.02 \\
NGC~6266 &  12.1 $\pm$ 2.4 $^a$ &   -1.18 $\pm$ 0.07  &   -1.20   &  &  10.60 $\pm$    0.10 &   -1.22 $\pm$    0.01 \\
NGC~6284 &  11.4 $\pm$ 2.3 $^a$ &   -1.31 $\pm$ 0.09  &   -1.27   &  &  10.21 $\pm$    0.12 &   -1.33 $\pm$    0.01 \\
        &                      &                     &           &  &  10.39 $\pm$    0.48 &   -1.28 $\pm$    0.02 \\
NGC~6304 &  13.5 $\pm$ 2.8 $^b$ &   -0.37 $\pm$ 0.07  &   -0.66   &  &  14.69 $\pm$    3.43 &   -0.63 $\pm$    0.06 \\
NGC~6316 &                      &   -0.36 $\pm$ 0.14  &   -0.90   &  &  14.83 $\pm$    0.93 &   -0.86 $\pm$    0.02 \\
        &                      &                     &           &  &  14.46 $\pm$    1.30 &   -0.86 $\pm$    0.03 \\
NGC~6333 &                      &   -1.79 $\pm$ 0.09  &   -1.65   &  &  10.49 $\pm$    2.17 &   -2.03 $\pm$    0.03 \\
NGC~6342 &  12.3 $\pm$ 2.5 $^a$ &   -0.49 $\pm$ 0.14  &   -1.01   &  &  13.13 $\pm$    0.47 &   -0.93 $\pm$    0.01 \\
        &                      &                     &           &  &  12.95 $\pm$    0.85 &   -0.84 $\pm$    0.02 \\
NGC~6352 &  12.6 $\pm$ 2.6 $^b$ &   -0.62 $\pm$ 0.05  &   -0.70   &  &  12.24 $\pm$    0.35 &   -0.70 $\pm$    0.01 \\
NGC~6356 &                      &   -0.35 $\pm$ 0.14  &   -0.74   &  &  14.26 $\pm$    2.10 &   -0.76 $\pm$    0.04 \\
NGC~6362 &  12.1 $\pm$ 2.4 $^a$ &   -1.07 $\pm$ 0.05  &   -1.17   &  &  14.53 $\pm$    0.21 &   -1.22 $\pm$    0.01 \\
NGC~6388 &  12.0 $\pm$ 2.6 $^b$ &   -0.45 $\pm$ 0.04  &   -0.68   &  &   6.57 $\pm$    1.10 &   -0.62 $\pm$    0.03 \\
NGC~6441 &  11.2 $\pm$ 2.4 $^b$ &   -0.44 $\pm$ 0.07  &   -0.65   &  &   6.09 $\pm$    1.80 &   -0.55 $\pm$    0.04 \\
        &                      &                     &           &  &   5.89 $\pm$    0.67 &   -0.54 $\pm$    0.02 \\
NGC~6522 &                      &   -1.45 $\pm$ 0.08  &   -1.39   &  &   8.97 $\pm$    0.43 &   -1.24 $\pm$    0.02 \\
NGC~6528 &  12.6 $\pm$ 2.5 $^c$ &    0.07 $\pm$ 0.08  &   -0.10   &  &  13.60 $\pm$    1.42 &   -0.29 $\pm$    0.03 \\
        &                      &                     &           &  &  12.39 $\pm$    1.95 &   -0.26 $\pm$    0.04 \\
NGC~6544 &  10.6 $\pm$ 2.3 $^a$ &   -1.47 $\pm$ 0.07  &   -1.38   &  &  11.57 $\pm$    0.91 &   -1.34 $\pm$    0.02 \\
NGC~6553 &  13.0 $\pm$ 2.5 $^d$ &   -0.16 $\pm$ 0.06  &   -0.20   &  &  10.35 $\pm$    4.23 &   -0.29 $\pm$    0.08 \\
NGC~6569 &                      &   -0.72 $\pm$ 0.14  &   -1.08   &  &  14.98 $\pm$    0.15 &   -1.06 $\pm$    0.01 \\
NGC~6624 &  12.5 $\pm$ 2.6 $^b$ &   -0.42 $\pm$ 0.07  &   -0.70   &  &   9.21 $\pm$    1.53 &   -0.70 $\pm$    0.04 \\
        &                      &                     &           &  &  11.21 $\pm$    1.71 &   -0.74 $\pm$    0.04 \\
NGC~6626 &                      &   -1.46 $\pm$ 0.09  &   -1.21   &  &  10.39 $\pm$    0.68 &   -1.37 $\pm$    0.02 \\
NGC~6637 &  11.9 $\pm$ 2.6 $^a$ &   -0.59 $\pm$ 0.07  &   -0.78   &  &  13.38 $\pm$    1.34 &   -0.86 $\pm$    0.03 \\
NGC~6638 &                      &   -0.99 $\pm$ 0.07  &   -1.08   &  &  12.96 $\pm$    1.21 &   -1.00 $\pm$    0.02 \\
NGC~6652 &  12.1 $\pm$ 2.5 $^a$ &   -0.76 $\pm$ 0.14  &   -1.10   &  &  11.72 $\pm$    0.28 &   -0.95 $\pm$    0.01 \\
        &                      &                     &           &  &  10.12 $\pm$    0.15 &   -1.01 $\pm$    0.01 \\
NGC~6723 &  12.7 $\pm$ 2.9 $^a$ &   -1.10 $\pm$ 0.07  &   -1.14   &  &  12.78 $\pm$    0.15 &   -1.34 $\pm$    0.01 \\
NGC~6752 &  13.5 $\pm$ 2.9 $^a$ &   -1.55 $\pm$ 0.01  &   -1.57   &  &   9.37 $\pm$    3.75 &   -1.86 $\pm$    0.07 \\
NGC~7078 &  12.5 $\pm$ 2.6 $^a$ &   -2.33 $\pm$ 0.02  &    --     &  &   6.82 $\pm$    1.43 &   -2.32 $\pm$    0.01 \\
        &                      &                     &           &  &   6.98 $\pm$    1.19 &   -2.32 $\pm$    0.01 \\
NGC~7089 &  11.9 $\pm$ 2.7 $^a$ &   -1.66 $\pm$ 0.07  &   -1.49   &  &  10.36 $\pm$    2.89 &   -1.82 $\pm$    0.06 \\
\end{longtable}
\begin{minipage}{.90\hsize}
{\bf Notes:} For GCs observed more than once, each spectrum was fitted separately. 
Our results correspond to the average and 
1$\sigma$ values of the MC simulations (see text in \S 3).
{\bf References:} 
(a)~\citet{deangeli+05}; (b)~\citet{marin-franch+09}; (c)~\citet{momany+03}; (d)~\citet{beaulieu+01}. Ages from references (a) and (b) were
converted to absolute values adopting an age of 13~$\pm$~2.5 Gyr for 47~Tuc \citep{zoccali+01}, and errors are propagated from the original
references. {\bf Column 3 and 4}: [Fe/H] from \citet{carretta+09} and \citet{schiavon+05}, respectively. {\bf Columns 5 and 6:} ages and [Fe/H] obtained in this work, respectively.
\end{minipage}
}


\longtab{4}{
\begin{longtable}{l r r l c}
\caption[Literature data]{\label{tab_lit_m31}
Ages and metallicities from literature, for our sample of M31 GCs.}\\
\hline
Cluster & Age    & [Fe/H] & Method & Reference    \\
        &  (Gyr) & (dex)  &        &               \\  
\hline
\endfirsthead

\caption{continued.}\\
\hline
Cluster & Age    & [Fe/H] & Method & Reference    \\
        &  (Gyr) & (dex)  &        &               \\  
\hline
\endhead

\multicolumn{5}{r}{{Continued on next page}} \\ \hline
\endfoot

\hline
\endlastfoot
B126 &     3.40 $\pm$    1.00  &    -1.43 $\pm$    0.42 & Spectral indices &   a \\
     &     5.50 $\pm$    3.50  &    -1.56 $\pm$    0.24 & Spectral indices &   b \\
     &     7.20 $\pm$    3.10  &    -1.39 $\pm$    0.28 & Spectral indices &   c \\
B134 &     9.10 $\pm$    2.20  &    -0.99 $\pm$    0.48 & Spectral indices &   a \\
     &    11.90 $\pm$    1.90  &    -1.00 $\pm$    0.38 & Spectral indices &   b \\
     &    11.30 $\pm$    1.80  &     0.46 $\pm$    0.11 & Spectral indices &   c \\
     &     5.50 $\pm$    2.28  &    -0.64 $\pm$    0.08 & Photometry       &   d \\
B158 &     9.90 $\pm$    2.70  &    -1.01 $\pm$    0.24 & Spectral indices &   a \\
     &    12.10 $\pm$    0.90  &    -0.84 $\pm$    0.18 & Spectral indices &     b \\
     &    11.30 $\pm$    1.50  &    -0.70 $\pm$    0.08 & Spectral indices &   c \\
     &    20.00 $\pm$    5.57  &    -1.02 $\pm$    0.02 & Photometry       &   d \\
B163 &    10.60 $\pm$    3.70  &    -0.25 $\pm$    0.27 & Spectral indices &     a \\
     &    10.20 $\pm$    4.80  &    -0.24 $\pm$    0.22 & Spectral indices &     b \\
     &     7.70 $\pm$    1.00  &    -0.10 $\pm$    0.06 & Spectral indices &   c \\
     &    11.75 $\pm$    1.60  &    -0.36 $\pm$    0.27 & Photometry       &   d \\
B222 &     0.70 $\pm$    0.70  &     0.30 $\pm$    0.60 & Spectral indices &     e \\
     &     1.10 $\pm$    0.40  &    -0.65 $\pm$    0.18 & Spectral indices &   c \\
     &     7.75 $\pm$    1.46  &    -0.93 $\pm$    0.95 & Photometry       &   d \\
B225 &     8.40 $\pm$    1.90  &    -0.49 $\pm$    0.14 & Spectral indices &     a \\
     &     9.10 $\pm$    1.20  &    -0.39 $\pm$    0.18 & Spectral indices &     b \\
     &     9.90 $\pm$    1.20  &    -0.45 $\pm$    0.05 & Spectral indices &   c \\
     &    15.50 $\pm$    4.89  &    -0.67 $\pm$    0.12 & Photometry       &   d \\
B234 &    10.30 $\pm$    3.20  &    -0.86 $\pm$    0.18 & Spectral indices &     a \\
     &    11.70 $\pm$    5.70  &    -0.84 $\pm$    0.19 & Spectral indices &     b \\
     &    11.40 $\pm$    2.00  &    -0.54 $\pm$    0.10 & Spectral indices &   c \\ 
B292 &     2.70 $\pm$    1.20  &    -1.91 $\pm$    0.58 & Spectral indices &     a \\
     &     5.90 $\pm$    3.40  &    -1.70 $\pm$    0.20 & Spectral indices &      b \\
     &     9.20 $\pm$    3.30  &    -1.63 $\pm$    0.49 & Spectral indices &   c \\ 
     &     1.00 $\pm$    0.10  &    -1.42 $\pm$    0.16 & Photometry &   f \\ 
B301 &     3.20 $\pm$    1.80  &    -1.29 $\pm$    0.31 & Spectral indices &     a \\
     &     6.20 $\pm$    1.10  &    -1.27 $\pm$    0.55 & Spectral indices &     b \\
     &     4.10 $\pm$    3.70  &    -0.50 $\pm$    0.26 & Spectral indices &   c \\ 
     &                         &    -0.76 $\pm$    0.25 & Photometry &   f \\ 
B304 &    12.90 $\pm$    4.70  &    -1.55 $\pm$    0.65 & Spectral indices &     a \\
     &    11.90 $\pm$    3.90  &    -1.45 $\pm$    0.43 & Spectral indices &     b \\
     &    13.20 $\pm$    3.10  &    -1.16 $\pm$    0.46 & Spectral indices &   c \\ 
     &                         &    -1.32 $\pm$    0.22 & Photometry &   f \\ 
B305 &     0.40 $\pm$    0.10  &    -0.90 $\pm$    0.61 & Photometry &   f \\ 
B307 &     1.61 $\pm$    0.10  &    -0.41 $\pm$    0.36 & Photometry &   f \\ 
B310 &    19.50 $\pm$    2.10  &    -2.06 $\pm$    0.51 & Spectral indices &     a \\
     &    11.40 $\pm$    3.50  &    -1.69 $\pm$    0.47 & Spectral indices &     b \\
     &    10.80 $\pm$    3.10  &    -1.24 $\pm$    0.52 & Spectral indices &   c \\ 
     &                         &    -1.43 $\pm$    0.28 & Photometry &   f \\ 
B313 &     8.80 $\pm$    1.20  &    -1.05 $\pm$    0.35 & Spectral indices &      a \\
     &    11.70 $\pm$    0.90  &    -0.89 $\pm$    0.50 & Spectral indices &     b \\
     &    11.20 $\pm$    1.20  &    -0.79 $\pm$    0.32 & Spectral indices &   c \\ 
     &     7.28 $\pm$    0.70  &    -1.09 $\pm$    0.10 & Photometry &   f \\ 
B314 &     0.50 $\pm$    0.60  &     0.35 $\pm$    0.25 & Spectral indices &     e \\ 
     &     1.00 $\pm$    0.10  &    -0.35 $\pm$    0.22 & Spectral indices &   c \\ 
B316 &     1.06 $\pm$    0.10  &    -1.47 $\pm$    0.23 & Photometry &   f \\ 
B321 &     0.30 $\pm$    0.30  &     0.00 $\pm$    0.60 & Spectral indices &     e \\ 
     &     1.00 $\pm$    0.10  &    -1.98 $\pm$    0.30 & Spectral indices &   c \\ 
B322 &     0.10 $\pm$    0.50  &    -0.20 $\pm$    0.20 & Spectral indices &     e \\ 
     &     2.30 $\pm$    0.70  &    -1.99 $\pm$    0.43 & Spectral indices &   c \\ 
B324 &     0.90 $\pm$    0.20  &    -0.05 $\pm$    0.40 & Spectral indices &     e \\ 
     &     1.00 $\pm$    0.10  &    -0.58 $\pm$    0.20 & Spectral indices &   c \\ 
B327 &     0.10 $\pm$    0.90  &     0.30 $\pm$    0.75 & Spectral indices &     e \\ 
     &     5.40 $\pm$    1.40  &    -1.97 $\pm$    0.34 & Spectral indices &   c \\  
B328 &    16.80 $\pm$    5.20  &    -2.13 $\pm$    0.35 & Spectral indices &     a \\
     &    13.30 $\pm$    1.30  &    -1.69 $\pm$    0.35 & Spectral indices &     b \\
     &    11.00 $\pm$    2.40  &    -2.29 $\pm$    0.42 & Spectral indices &   c \\ 
B337 &     2.60 $\pm$    1.90  &    -1.18 $\pm$    0.19 & Spectral indices &     a \\
     &     6.60 $\pm$    3.20  &    -1.19 $\pm$    0.25 & Spectral indices &     b \\
     &     4.90 $\pm$    2.90  &    -0.59 $\pm$    0.11 & Spectral indices &   c \\ 
     &     2.03 $\pm$    0.10  &    -1.09 $\pm$    0.32 & Photometry &   f \\
B347 &     8.70 $\pm$    5.70  &    -2.49 $\pm$    0.21 & Spectral indices &     a \\
     &     9.90 $\pm$    4.00  &    -1.93 $\pm$    0.51 & Spectral indices &     b \\
     &    10.20 $\pm$    2.40  &    -2.61 $\pm$    0.37 & Spectral indices &   c \\
     &     2.53 $\pm$    0.15  &    -1.71 $\pm$    0.03 & Photometry &   f \\
B350 &     9.80 $\pm$    2.50  &    -1.98 $\pm$    0.49 & Spectral indices &     a \\
     &    12.40 $\pm$    4.70  &    -1.71 $\pm$    0.33 & Spectral indices &     b \\
     &     9.30 $\pm$    2.30  &    -1.81 $\pm$    0.37 & Spectral indices &   c \\
     &     1.99 $\pm$    0.10  &    -1.47 $\pm$    0.17 & Photometry &   f \\
B354 &     5.24 $\pm$    0.65  &    -1.46 $\pm$    0.38 & Photometry &   f \\
B365 &     9.20 $\pm$    3.10  &    -1.59 $\pm$    0.44 & Spectral indices &     a \\
     &    10.40 $\pm$    2.20  &    -1.42 $\pm$    0.51 & Spectral indices &     b \\
     &     6.60 $\pm$    3.10  &    -1.03 $\pm$    0.20 & Spectral indices &   c \\  
     &     1.73 $\pm$    0.10  &    -1.78 $\pm$    0.19 & Photometry &   f \\
B380 &     0.45 $\pm$    0.13  &     0.15 $\pm$    0.10 & Spectral indices &     e \\
     &     1.00 $\pm$    0.10  &    -1.69 $\pm$    0.13 & Spectral indices &   c \\
B383 &    10.30 $\pm$    2.20  &    -0.68 $\pm$    0.18 & Spectral indices &     a \\
     &    10.60 $\pm$    1.30  &    -0.68 $\pm$    0.31 & Spectral indices &     b \\
     &    11.30 $\pm$    2.40  &    -0.59 $\pm$    0.08 & Spectral indices &   c \\
     &    13.99 $\pm$    1.05  &    -0.48 $\pm$    0.20 & Photometry &   f \\
B393 &    10.00 $\pm$    1.80  &     0.00 $\pm$    0.00 & Spectral indices &     a \\
     &    11.90 $\pm$    1.30  &    -0.80 $\pm$    0.47 & Spectral indices &     b \\
     &    11.20 $\pm$    1.40  &    -0.60 $\pm$    0.14 & Spectral indices &   c \\
     &     6.76 $\pm$    1.10  &    -1.41 $\pm$    0.05 & Photometry  & f      \\
B398 &    12.10 $\pm$    1.50  &    -0.71 $\pm$    0.41 & Spectral indices &     a \\
     &    14.70 $\pm$    3.70  &    -0.52 $\pm$    0.28 & Spectral indices &     b \\
     &    15.00 $\pm$    2.40  &    -0.54 $\pm$    0.10 & Spectral indices &   c \\
B401 &     9.20 $\pm$    5.80  &    -2.49 $\pm$    0.49 & Spectral indices &     a \\
     &    15.00 $\pm$    5.10  &    -2.38 $\pm$    0.24 & Spectral indices &     b \\
     &    10.20 $\pm$    2.40  &    -2.38 $\pm$    0.32 & Spectral indices &   c \\
     &     3.49 $\pm$    0.40  &    -1.75 $\pm$    0.29 & Photometry  &     f  \\
MGC1 &     7.10 $\pm$    3.00  &    -1.37 $\pm$    0.15 & Spectral indices &   g \\
MCGC5 &    10.00 $\pm$    3.00  &    -1.33 $\pm$    0.12 & Spectral indices &   g \\
MCGC10 &   12.60 $\pm$    3.00  &    -1.73 $\pm$    0.20 & Spectral indices &   g \\
NB16 &     2.00 $\pm$    1.40  &    -1.40 $\pm$    0.18 & Spectral indices &     a \\
     &     4.60 $\pm$    0.80  &    -1.33 $\pm$    0.24 & Spectral indices &     b \\
     &     7.20 $\pm$    3.50  &    -1.11 $\pm$    0.15 & Spectral indices &   c \\
NB89 &     8.30 $\pm$    2.60  &    -0.78 $\pm$    0.14 & Spectral indices &     a \\
     &    14.40 $\pm$    2.70  &    -0.84 $\pm$    0.15 & Spectral indices &     b \\
     &     6.80 $\pm$    1.70  &    -0.58 $\pm$    0.08 & Spectral indices &   c \\
\hline
\end{longtable}
\begin{minipage}{.90\hsize}
{\bf References for column 5:}
(a) \citet{beasley+05} \citep[models by][]{BC03};
(b) \citet{beasley+05} \citep[models by][]{tmb03,tmk04};
(c) \citet{puzia+05};
(d) \citet{fan+06};
(e) \citet{beasley+04};
(f) \citet{wang+10}, and; 
(g) \citet{abrito+09}. Regarding the references (a), (b) and (c), we converted their [Z/H] to [Fe/H] through equation 4 in \citet{tmb03}.
{\bf Notes:} We did not find in the listed literature the parameters for two clusters in our sample, B302 and B331. 
\end{minipage}
}

\end{document}